\documentclass[a4paper,11pt]{scrartcl}

\usepackage[top=3cm,left=3cm,right=3cm,bottom=3cm]{geometry}
\usepackage[intlimits]{amsmath}
\usepackage{amsfonts,amssymb,amsthm}
\usepackage{hyperref}
\usepackage{enumerate}
\usepackage{braket}
\usepackage{graphbox}
\usepackage{booktabs}
\usepackage[section]{placeins}
\usepackage{caption}
\usepackage{subcaption}
\usepackage{color}
\usepackage{todonotes}

\hypersetup{pdfstartview=FitH}

\begin{document}

\title{Decoherence on Staggered Quantum Walks}
\author{Raqueline A. M. Santos $^\dagger$ and Franklin de L. Marquezino $^\ddagger$}

\date{\small
$^\dagger$ Center for Quantum Computer Science, Faculty of Computing,\\ University of Latvia, Latvia, rsantos@lu.lv\\
$^\ddagger$ Federal University of Rio de Janeiro, Brazil, franklin@cos.ufrj.br}
\maketitle

\begin{abstract}
Decoherence phenomenon has been widely studied in different types of quantum walks. In this work we show how to model decoherence inspired by percolation on staggered quantum walks. Two models of unitary noise are described: breaking polygons and breaking vertices. The evolution operators subject to these noises are obtained and the equivalence to the coined quantum walk model is presented. Further, we numerically analyze the effect of these decoherence models on the two-dimensional grid of $4$-cliques.  We examine how these perturbations affect the quantum walk based search algorithm in this graph and how expanding the tessellations intersection can make it more robust against decoherence.
\end{abstract}

\section{Introduction}

As the quantum counterparts of classical random walks, quantum walks have been defined in different ways in both discrete and continuous time~\cite{Aharonov:1993,Farhi:1998,Szegedy:2004}. They are a powerful tool in the development of efficient quantum algorithms~\cite{Shenvi:2003,Ambainis:2004, Childs:2004} and in the simulation of complex physical systems~\cite{Childs:2009,Berry:2012}.
More recently, Portugal \emph{et al.}~\cite{Portugal:2015} described a model of quantum walks on arbitrary graphs called the staggered quantum walk (SQW) model. This model is defined by partitioning the graph into tessellations. Each tessellation is a partition of the vertex set into polygons (or cliques). The relation of the SQW with other quantum walk models was studied in~\cite{Portugal:2015,Portugal:2016,Coutinho:2018, Konno:2018, Sadowski:2019}. It has been also applied in the development of quantum algorithms~\cite{Portugal:2017a,Chagas:2018,Portugal:2018}, and in physical implementations~\cite{Jalil:2017}.

It is known that when implementing quantum systems we can face decoherence problems.  Quantum walks implementations are also affected by it. It is important to understand, for example, when the classical behavior emerges and the quantum effect disappears. The quantum coherence is affected by the influence of random events which can be modeled in different ways. Refer to \cite{Kendon:2007} for different methods of simulating decoherence in quantum walks.
Decoherence inspired by percolation involves randomly removing or creating vertices or edges in the graph. This type of decoherence was analyzed in many research papers, using the discrete-time  coined model~\cite{Romanelli:2005, Oliveira:2006,Kollar:2012,Lovett:2018,Tude:2021},  Szegedy's model~\cite{Santos:2014}, and the continuous-time quantum walk model~\cite{Xu:2008,Darazs:2013,Benedetti:2019}.

Therefore, it is crucial to understand how decoherence affects the SQW model as well.
In this paper, we analyze decoherence inspired by percolation on the SQW model. We describe two models of unitary noise: one by breaking vertices of the graph and another inherent to the SQW, by breaking polygons. The latter is equivalent to breaking some edges of the graph, such that no additional tessellations are necessary. We describe how to obtain the evolution operators subject to these noises and we show the equivalence to the coined quantum walk, when the SQW can be obtained from the coined model. Moreover, we numerically study how decoherence affects the SQW on the two-dimensional grid of $4$-cliques. We analyze the behavior of the quantum walk in terms of the displacement of the walker during time and we examine how the perturbations affect the search for a marked vertex in this graph. Additionally, we observe how the search algorithm can be more robust against decoherence when we expand the tessellations intersection of the graph.

The paper is organized as follows. In Sec.~\ref{sec:sqw}, we describe the SQW model. In Sec.~\ref{sec:decoherence}, we introduce the models of decoherence inspired by percolation and its equivalence to the coined model. In Sec.~\ref{sec:simulations}, we numerically analyze the effect of decoherence on the SQW on the two-dimensional grid of $4$-cliques. Conclusions and further discussions are drawn in Sec.~\ref{sec:conc}. 

\section{The Staggered Quantum Walk model}\label{sec:sqw}

The SQW model is defined by obtaining a \textit{tessellation cover} of the graph. A \textit{tessellation}~$\mathcal{T}$ of a graph $G$ is a partition of $G$ into cliques. A \textit{clique} of a graph $G$ is a complete subgraph of $G$. A clique of size $d$ is called a $d$-clique. Each clique in a tessellation is called a \textit{polygon} or a \textit{cell}. We say that an edge \textit{belongs to} a tessellation if both endpoints of the edge belong to the same polygon. We denote by~$\mathcal{E}(\mathcal{T})$ the set of edges belonging to~$\mathcal{T}$. Notice that a tessellation does not cover all the edges of a graph in general.
A \textit{tessellation cover} is a set $\{ \mathcal{T}_1, \cdots, \mathcal{T}_n \}$ of tessellations of a graph $G$, such that the union $\bigcup_{i=1}^{k} \mathcal{E}(\mathcal{T}_i)$ is the edge set of $G$.
The minimum number of tessellations required to cover a graph $G$ is the \textit{tessellation number}, denoted by $T(G)$. If a graph $G$ is such that $T(G) \leq t$, for a fixed integer $t$, we say that $G$ is $t$-tessellable.

Let us define a SQW on a connected simple graph $G$. Suppose  $G$ is $\ell$-tessellable and the set $\{\mathcal{T}_1,\dots,\mathcal{T}_\ell\}$ is a tessellation cover of $G$. Let $V(G)$ denote the set of vertices of $G$ and $|V(G)| = N$. The $N$-dimensional Hilbert space $\mathcal{H}^{N}$ of the SQW is spanned by $\{\ket{v}:v\in V(G)\}$. Each basis state $\ket{v}$ is associated with a vertex $v$ of the graph.  Suppose  each tessellation $\mathcal{T}_k$ has $|\mathcal{T}_k|$ polygons denoted by $P_i^k$.
Each polygon is associated with a unit vector
\begin{equation}
    \ket{P_i^k} = \sum_{v\in V(P_i^k)}\alpha_{vik}\ket{v},
\end{equation}
with 
\begin{equation}
    \sum_{v\in V(P_i^k)}|\alpha_{vik}|^2 = 1.
\end{equation}
Each tessellation $\mathcal{T}_k$ is associated with a Hermitian operator
\begin{equation}
    U_{\mathcal{T}_k} = 2\sum_{j=1}^{|\mathcal{T}_k|}\ket{P_j^k}\bra{P_j^k}-I.
\end{equation}
The evolution operator of the SQW is given by
\begin{equation}
    U = U_{\mathcal{T}_{\ell}}\cdots U_{\mathcal{T}_2}U_{\mathcal{T}_1}.
    \label{eq:U}
\end{equation}

\begin{figure}[!htb]
\centering
    \begin{subfigure}[t]{0.48\textwidth}
         \centering
         \includegraphics[scale=0.55]{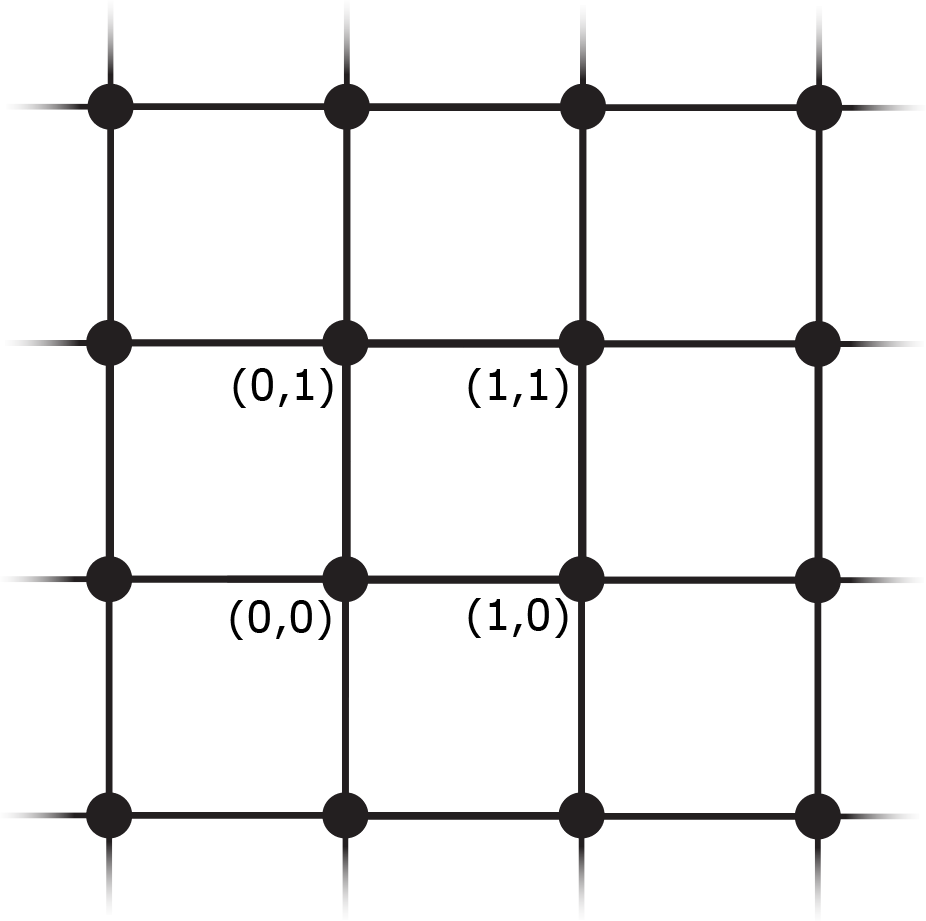}
         \caption{Two dimensional grid.}
         \label{fig:grid}
     \end{subfigure}\hfill
    \begin{subfigure}[t]{0.48\textwidth}
         \centering
         \includegraphics[scale=0.55]{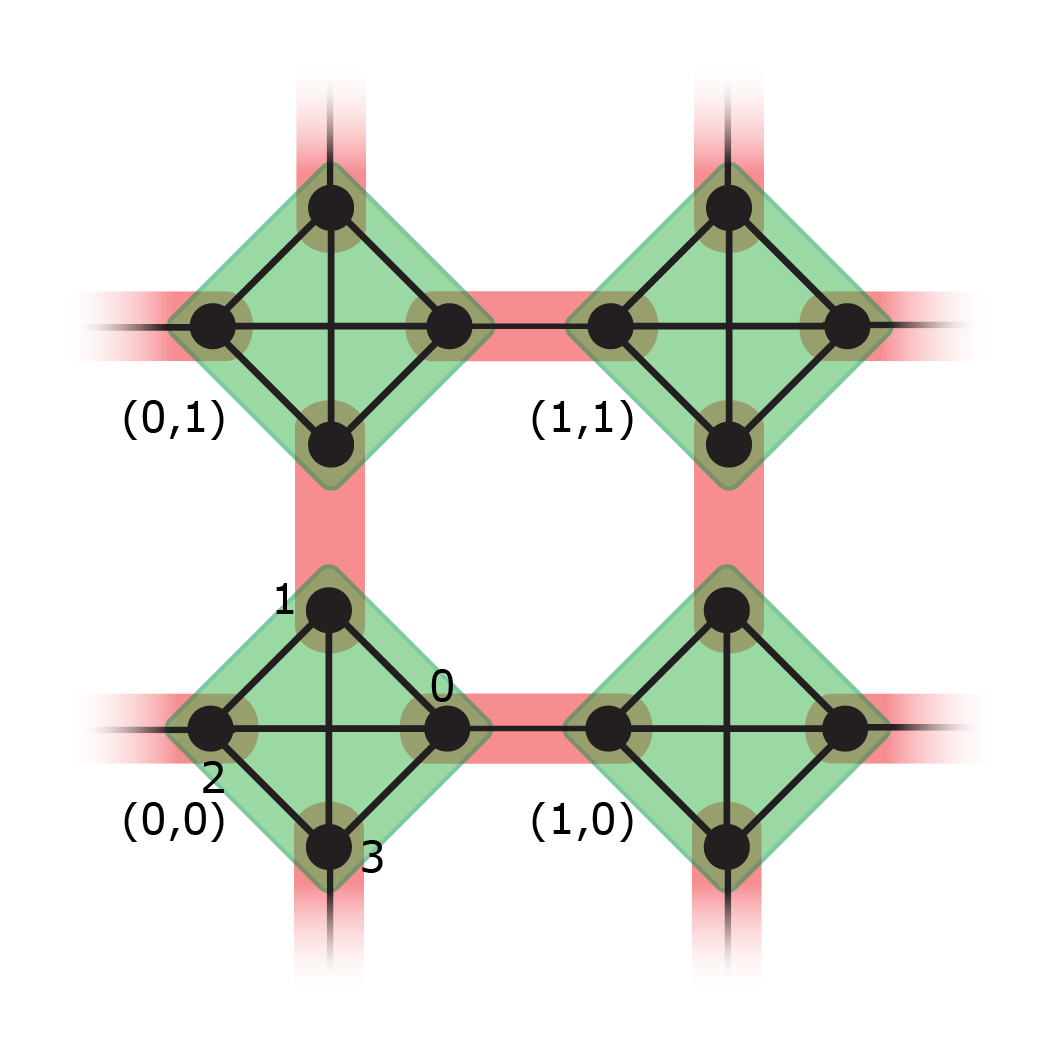}
         \caption{Two-dimensional grid of 4-cliques.}
         \label{fig:4-clique-grid}
     \end{subfigure}
     
     \begin{subfigure}[t]{0.48\textwidth}
         \centering
\includegraphics[scale=0.55]{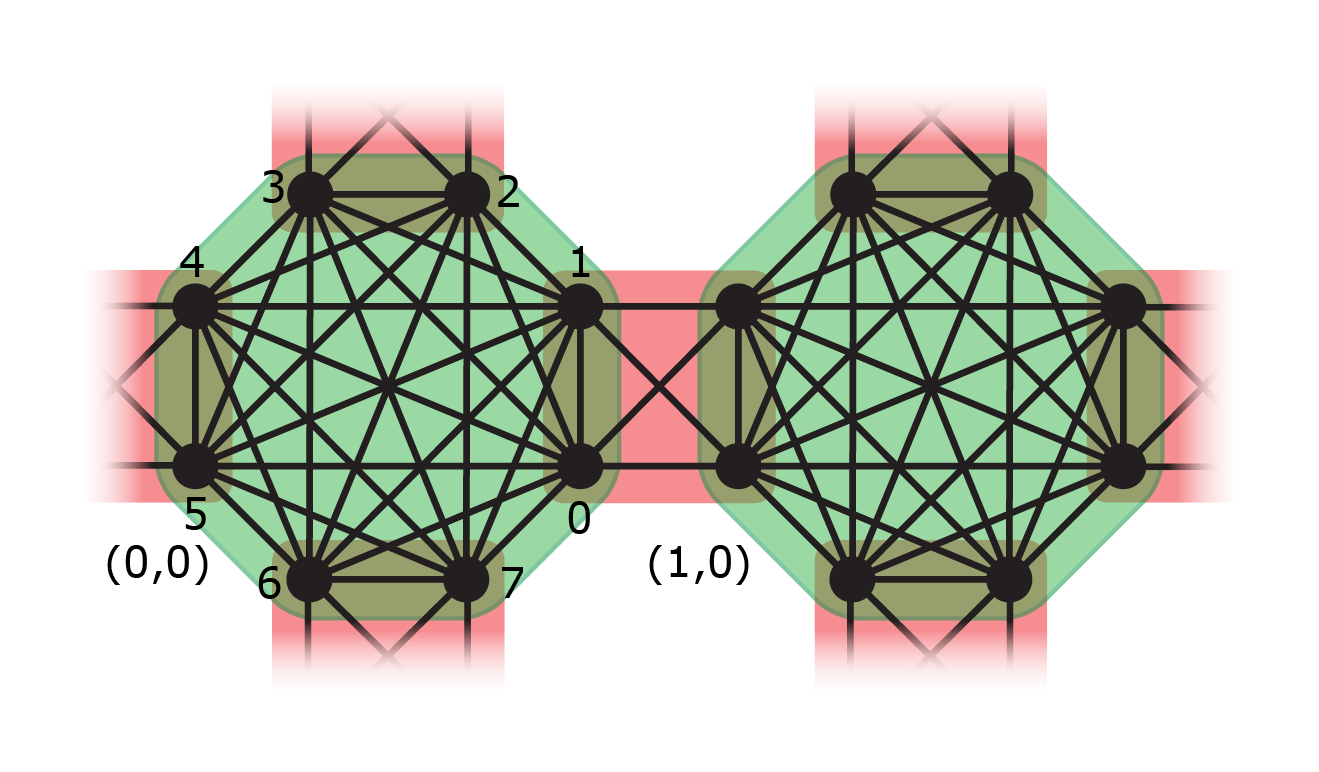}
\caption{Two-dimensional grid of 8-cliques.}
\label{fig:8-clique-grid}
\end{subfigure}
\begin{subfigure}[t]{0.48\textwidth}
         \centering
\includegraphics[scale=0.55]{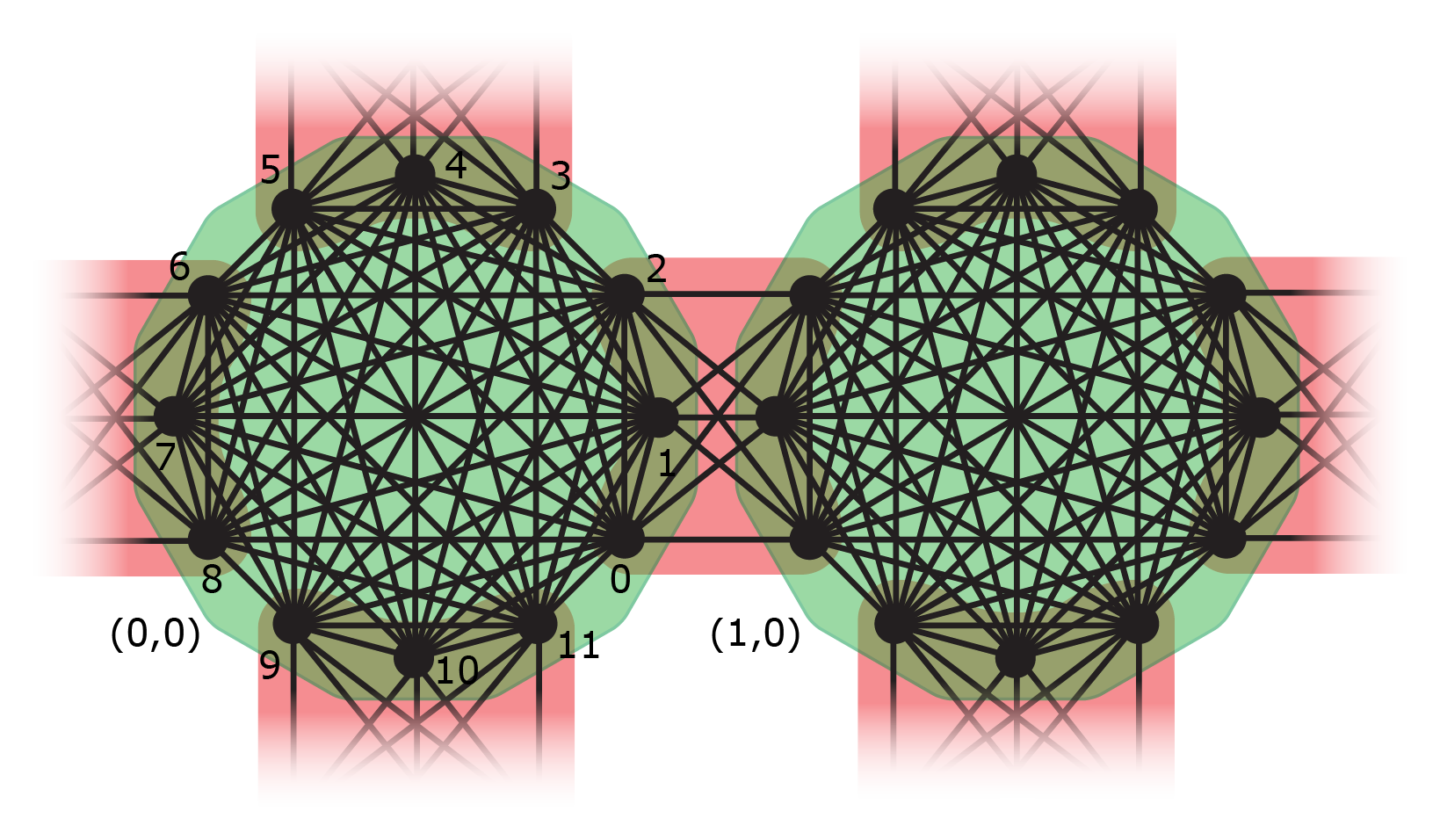}
\caption{Two-dimensional grid of 12-cliques.}
\label{fig:12-clique-grid}
\end{subfigure}
\caption{Examples of SQWs on the two-dimensional grids of $4q$-cliques with two tessellations, green and red. The SQW in (b) is the equivalent of the flip-flop coined quantum walk on the two-dimensional grid (a).}
\label{fig:sqws}
\end{figure}

For  example, let us consider a SQW on  two-dimensional grids of $4q$-cliques with periodic boundary conditions, as depicted in Figure~\ref{fig:sqws}. These graphs will be used throughout the paper.   Figures~\ref{fig:4-clique-grid},\ref{fig:8-clique-grid},\ref{fig:12-clique-grid} depict the cases where $q=1,2,3$, respectively.

The graph is composed by $n^2$ $4q$-cliques linked by $2n^2$ $2q$-cliques with a torus-like topology. The Hilbert space associated with the graph has dimension $4qn^2$.
The vectors associated with the green polygons are
\begin{equation}
\ket{{\alpha}_{xy}} = \frac{1}{2\sqrt{q}}\sum_{k=0}^{4q-1}\ket{x,y,k},
\label{eq:alpha}
\end{equation}
and the vectors associated with the red polygons are
\begin{eqnarray}
\ket{{\beta}_{xy}^{(0)}} &=& \frac{1}{\sqrt{2q}}\sum_{k=0}^{q-1}\left(\ket{x,y}\ket{k}+\ket{x+1,y}\ket{2q+k}\right),\label{eq:beta0}\\
\ket{{\beta}_{xy}^{(1)}} &=&\frac{1}{\sqrt{2q}}\sum_{k=0}^{q-1}\left(\ket{x,y}\ket{q+k}+\ket{x,y+1}\ket{3q+k}\right),\label{eq:beta1}
\end{eqnarray}
for $0\leq x,y \leq n-1$. The arithmetic with the labels of $\ket{x,y}$ is performed modulo $n$. 

The evolution operator is ${U} = {U}_1{U}_0$, where 
\begin{equation}
{U}_0 = 2\sum_{x,y = 0}^{n-1}\ket{{\alpha}_{xy}}\bra{{\alpha}_{xy}} - I,
\label{eq:U0}
\end{equation}
and
\begin{equation}
{U}_1 = 2\sum_{x,y = 0}^{n-1}\ket{{\beta}_{xy}^{(0)}}\bra{{\beta}_{xy}^{(0)}}+\ket{{\beta}_{xy}^{(1)}}\bra{{\beta}_{xy}^{(1)}} - I.
\label{eq:U1}
\end{equation}

\subsection{Equivalence to the coined model}

Consider a flip-flop coined quantum walk (FCQW) on a graph $G$. The Hilbert space of the walk is $\mathcal{H}^{2|E|}$, where $|E|$ is the number of edges of $G$. The basis state $\ket{u,v}$ represents the state of the walker at vertex $u$ pointing in the direction of vertex $v$. This walk is driven by the unitary operator $U = SC$, where $C$ is the coin operator that modifies the internal coin state of the walker and $S$ is the flip-flop shift which is responsible to move the walker between adjacent vertices, $S\ket{u,v} = \ket{v,u}$. In~\cite{Portugal:2013} you can find a detailed description of the coined quantum walk model. It is possible to obtain the equivalent SQW of the FCQW, as we briefly describe here. For more details, see~\cite{Portugal:2016}.

The SQW takes place in a space with dimension equal to the number of vertices of the graph, whereas the FCQW uses a bigger dimension space. In order to obtain the SQW equivalent of an FCQW, first we should create a new graph $G'$, so the dimensions will be equivalent. In this new graph, we convert each vertex $v \in V(G)$ into a $d(v)$-clique, where $d(v)$ is the degree of vertex $v$. 

\begin{table}[!htb]
\centering
\begin{tabular}{@{}ccc@{}}
\toprule
$d(v)$ & $v$ & $d(v)$-clique  \\
\midrule
\parbox[c][40pt][c]{0mm}{1}      & \includegraphics[align=c, scale=0.55]{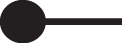}     & \includegraphics[align=c, scale=0.55]{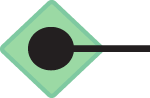}              \\
\parbox[c][40pt][c]{0mm}{2}      & \includegraphics[align=c, scale=0.55]{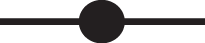}    & \includegraphics[align=c, scale=0.55]{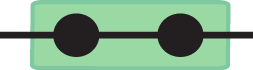}                \\
\parbox[c][40pt][c]{0mm}{3}      &  \includegraphics[align=c, scale=0.55]{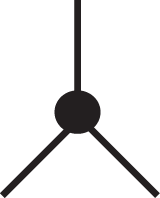}   & \includegraphics[align=c, scale=0.55]{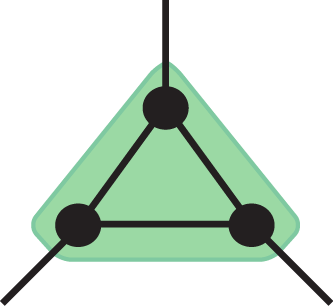}\\       \bottomrule         
\end{tabular}
\caption{Converting a vertex  $v$ of degree $d(v)$ into a $d(v)$-clique with associated green polygon belonging to the coin tessellation $\mathcal{T}_C$.}
\label{tb:convert}
\end{table}

\begin{figure}[!htb]
\centering
\begin{subfigure}[t]{0.3\textwidth}
         \centering
    
    \includegraphics[scale=0.55]{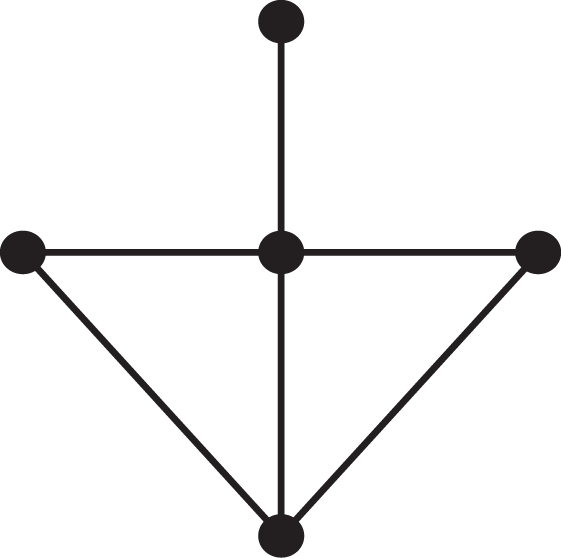}
    \caption{G.}
    \label{fig:convert_cliques}
\end{subfigure}
\hspace{5mm}
\begin{subfigure}[t]{0.3\textwidth}
         \centering
    
    \includegraphics[scale=0.55]{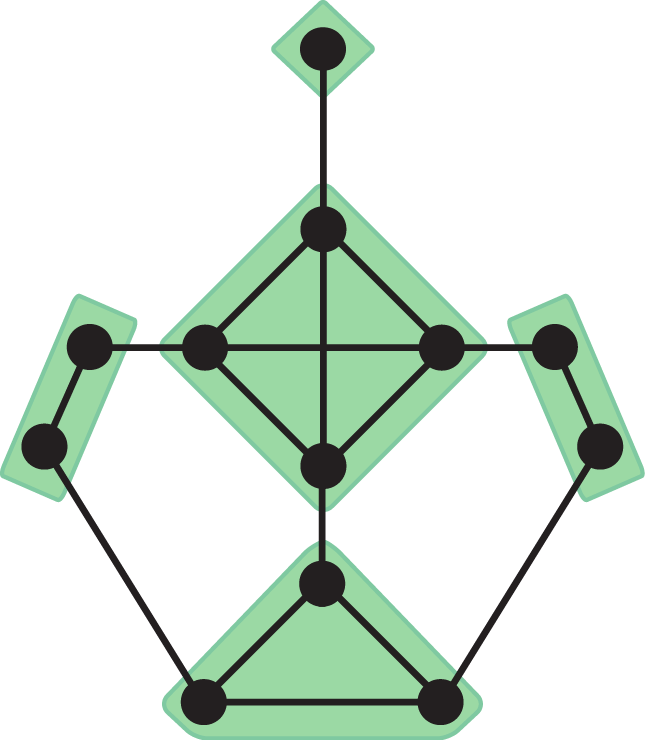}
    \caption{$G'$ with the coin tessellation in green.}
    \label{fig:graph_g}
\end{subfigure}
\hspace{5mm}
\begin{subfigure}[t]{0.3\textwidth}
         \centering
    
    \includegraphics[scale=0.55]{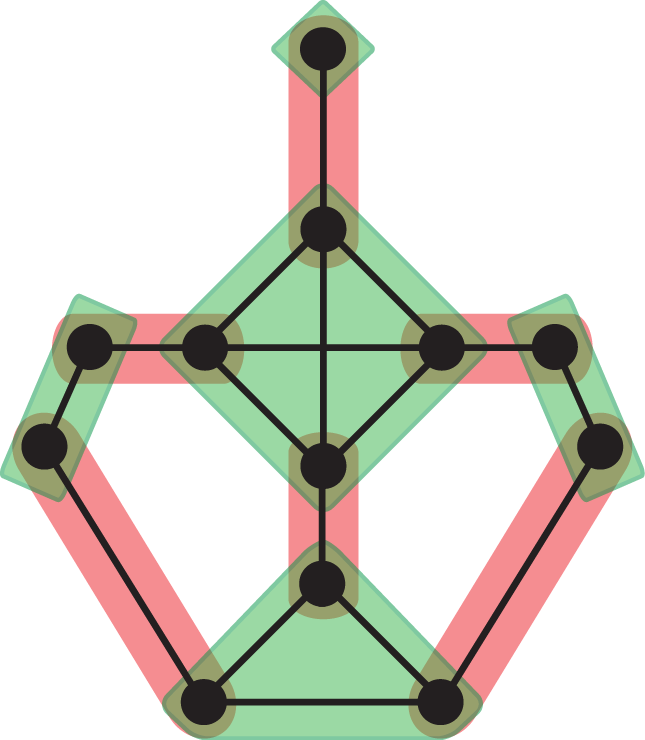}
    \caption{$G'$ with the coin tessellation in green and shift tessellation in red.}
    \label{fig:graph_g_prime}
\end{subfigure}

\caption{Graph $G'$ in (b) is obtained from $G$ in (a) by substituting each vertex $v\in V(G)$ by a $d(v)$-clique, which belongs to a green polygon of the coin tessellation $\mathcal{T}_C$. In (c) the shift tessellation $\mathcal{T}_S$ in red is obtained by each two connected vertices from different green polygons. }
\label{fig:convertSQW}
\end{figure}

Each created clique will belong to the coin tessellation $\mathcal{T}_C$, as we can see in the Table~\ref{tb:convert}. The $2$-clique formed by the connection between vertices of two different polygons in $\mathcal{T}_C$ will belong to the shift tessellation $\mathcal{T}_S$. 
See Figure~\ref{fig:convertSQW} for an example. The coin tessellation is represented in green and the shift tessellation in red. Notice also that Figure~\ref{fig:4-clique-grid} is the SQW equivalent of the FCQW on the two-dimensional grid (Figure~\ref{fig:grid}).

We can obtain exactly the shift and coin operators by the operators generated by each tessellation, that is, $U_{\mathcal{T}_C} \equiv C$ and $U_{\mathcal{T}_S} \equiv S$, as long as the coin operator is an orthogonal reflection. This condition is required in order to obtain the equivalent SQW. See~\cite{Portugal:2016} for a complete description of what is an orthogonal reflection and the whole equivalence process. The mostly used Grover and Hadamard coin operators are orthogonal reflections.

\subsection{Search with SQWs}

The implementation of spatial search on SQWs can be done by using partial tessellations. This is simply done by removing polygons from the tessellation. The vertices in the missing polygons will be the marked ones~\cite{Portugal:2015}.
\begin{figure}[!htb]
  \centering
\begin{subfigure}[t]{0.48\textwidth}
         \centering
    
    \includegraphics[scale=0.55]{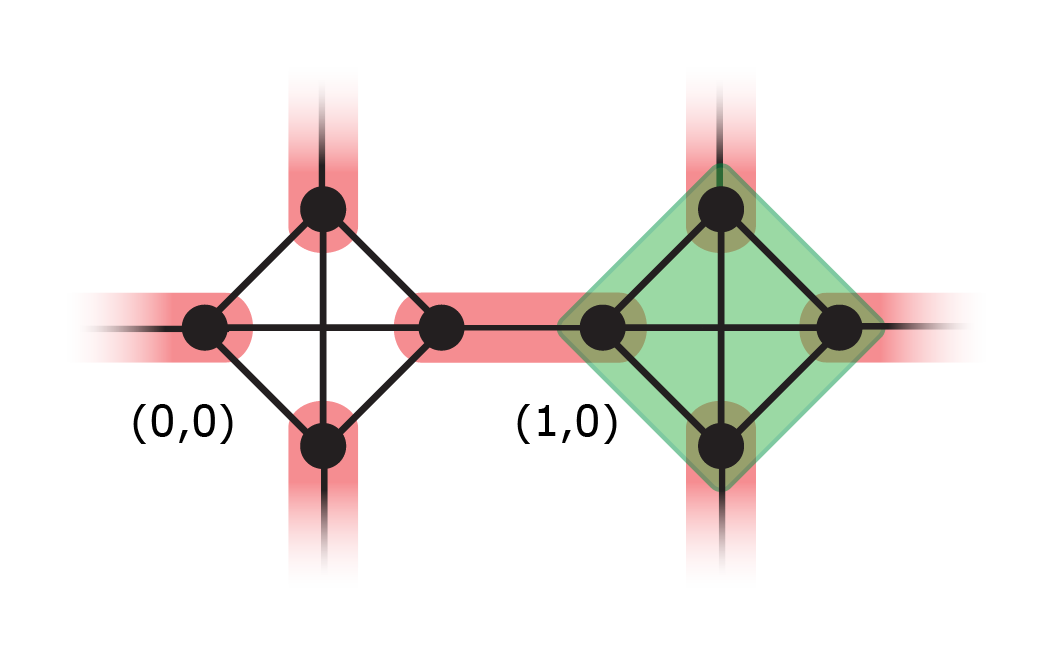}
    \caption{$q=1$.}
    \label{fig:sqw-search-q1}
\end{subfigure}
\begin{subfigure}[t]{0.48\textwidth}
         \centering
    
    \includegraphics[scale=0.55]{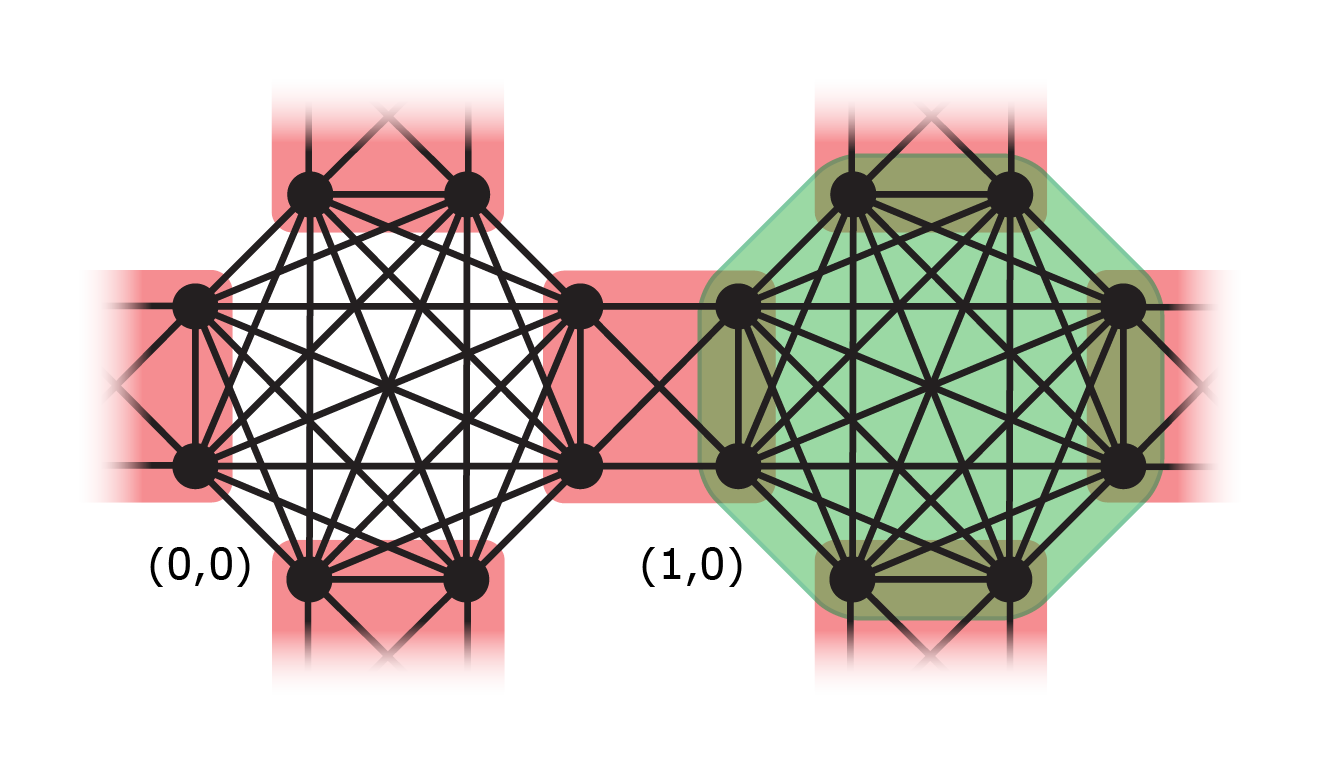}
    \caption{$q=2$.}
    \label{fig:sqw-search-q2}
\end{subfigure}
     \caption{Search for the $4q$-clique in position $(0,0)$. The green polygon containing the clique in position $(0,0)$ is removed and we have a SQW with a partial tessellation.}
     \label{fig:sqwsearch}
\end{figure}

Consider the example for the two-dimensional grid of $4q$-cliques presented earlier. Suppose that the $4q$-clique in position $(0,0)$ is the marked one, as depicted in Figure~\ref{fig:sqwsearch} for $q=1$ and $q=2$. We remove the polygon which induces the state $\ket{{\alpha}_{00}}$ in order to obtain the search operator ${U}' = {U}_1 {U}'_0$, where
$$
{U}'_0 = 2\sum_{\substack{x,y = 0\\(x,y)\neq (0,0)}}^{n-1}\ket{{{\alpha}}_{xy}}\bra{{{\alpha}}_{xy}} - I.
$$
The initial state $\ket{\psi(0)}$ is the uniform superposition of all vertices of the graph and the state at time $t$ is given by $\ket{\psi(t)} = {U}'^t\ket{\psi(0)}$.

Portugal~\cite{Portugal:2015} showed that searching for a marked vertex using FCQW with coin $-I$ on the marked vertex is equivalent to searching for a missing polygon in the equivalent SQW. This means that searching with SQW for a missing green polygon in the two-dimensional grid of 4-cliques (Figure~\ref{fig:sqw-search-q1}) is equivalent to search for a marked vertex in the two-dimensional grid using FCQW.

Recently, Santos~\cite{Santos:2019} showed that by a process called \emph{intersection expansion/reduction} we can add/remove vertices to some tessellations intersection of the SQW. The dynamics of the resulted SQW on the new graph will be equivalent to the dynamics on the original graph, if some assumptions are made for the vertices in the intersection. The two dimensional grids of $4q$-cliques can be obtained from each other by a process of tessellation intersection or reduction. Searching for a clique on the green polygons in any of these graphs is equivalent~\cite{Santos:2019}. 
The number of steps of the algorithm is $O(\sqrt{N\log N})$ and the success probability is $O(1/\log N)$, where $N=n^2$, and $n$ is the width of the grid. The total cost of the algorithm after applying the amplitude amplification method is $O(\sqrt{N}\log N)$.

\section{Decoherence inspired by percolation}
\label{sec:decoherence}

Decoherence inspired by percolation allows removing of vertices and/or edges in the graph.
What happens when we break edges and/or vertices in the SQW model? Is it possible to ``break'' polygons as well? 
The staggered model is described by a graph tessellation cover in order to obtain the evolution operator. The tessellations and its polygons follow some strict properties, as we mentioned before. In order to model this kind of decoherence we should take care about the tessellation structure so that its properties are satisfied and the evolution continues to be unitary.
Following, we describe some possible models of decoherence inspired by percolation affecting the SQWs. The models are valid for general graphs despite the tessellation number.

\subsection{Breaking polygons}

\begin{figure}[!htb]
\centering

\begin{subfigure}[t]{0.47\textwidth}
         \centering
    
    \includegraphics[scale=0.55]{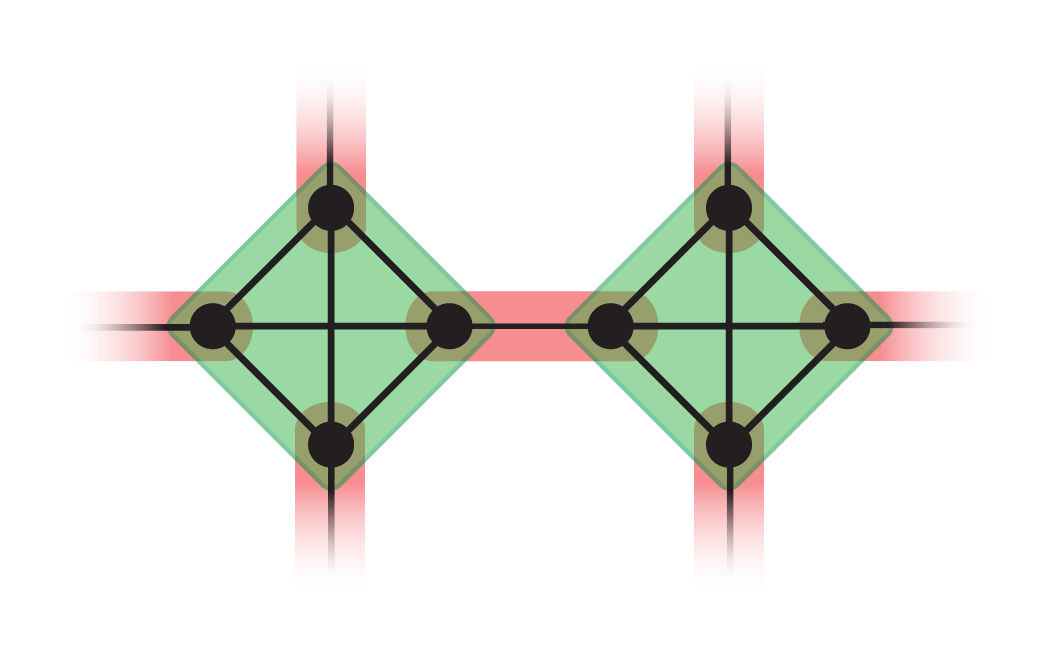}
    \caption{A two-tessellable graph with green and red tessellations.}
    \label{fig:sqw-grid}
\end{subfigure}
\hspace{5mm}
\begin{subfigure}[t]{0.47\textwidth}
         \centering
    
    \includegraphics[scale=0.55]{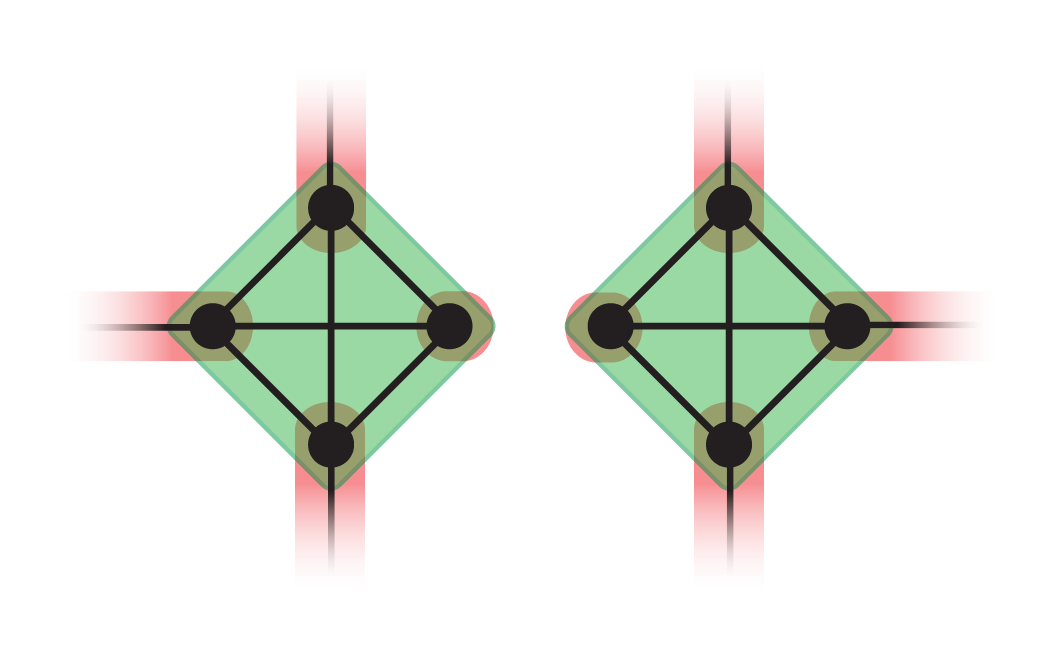}
    \caption{Breaking a red polygon.}
    \label{fig:break_poly_red}
\end{subfigure}
\begin{subfigure}[t]{0.3\textwidth}
         \centering
    
    \includegraphics[scale=0.55]{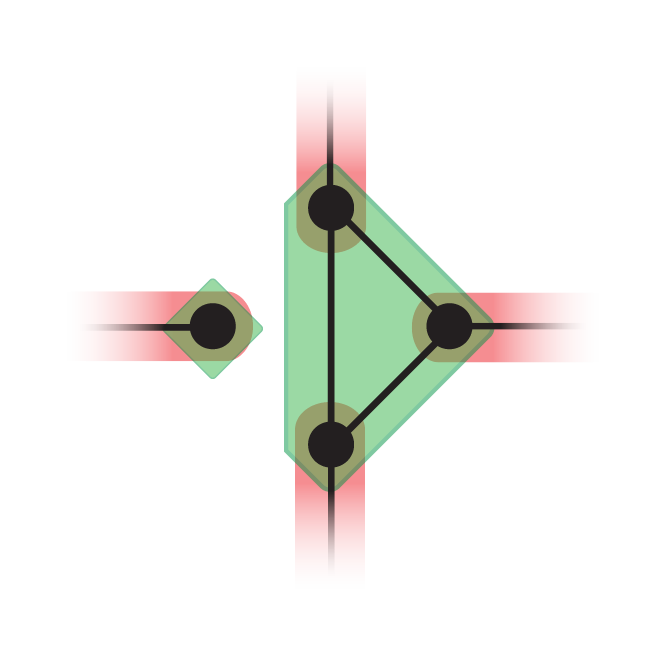}
    \caption{Breaking a green polygon into a 1-clique polygon and a 3-clique polygon.}
    \label{fig:break_poly_green_1}
\end{subfigure}
\hspace{5mm}
\begin{subfigure}[t]{0.3\textwidth}
         \centering
    
    \includegraphics[scale=0.55]{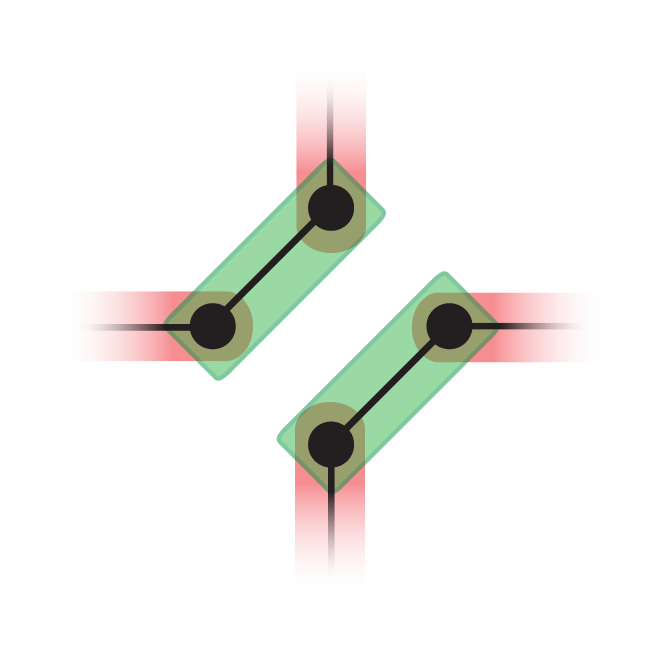}
    \caption{Breaking a green polygon into two 2-clique polygons.}
    \label{fig:break_poly_green_2}
\end{subfigure}
\hspace{5mm}
\begin{subfigure}[t]{0.3\textwidth}
         \centering
    
    \includegraphics[scale=0.55]{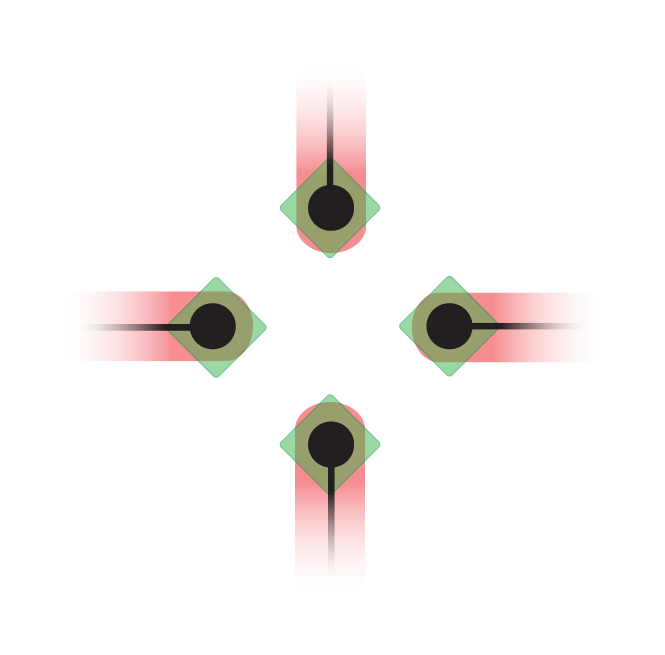}
    \caption{Breaking a green polygon into four 1-clique polygons.}
    \label{fig:break_poly_green_3}
\end{subfigure}

\caption{Example of breaking polygons on the SQW model.}
\label{fig:breaking_polygons}
\end{figure}

A polygon can be broken/divided into many polygons (as many as the number of vertices of the clique). Breaking a polygon is associated with breaking some edges of the clique, as we can see by the example in Figure~\ref{fig:breaking_polygons}. A polygon in the green tessellation can be broken into different ways (see Figures~\ref{fig:break_poly_green_1}-\ref{fig:break_poly_green_3}), whereas a red polygon can be broken only into two 1-clique polygons (see Figure~\ref{fig:break_poly_red}). That is because each red polygon is a 2-clique, in this case.

Let us assume that polygon $P_\ell^j$ was broken into $m \in \{2,\dots,|P_\ell^j|\}$  polygons. Each new polygon $P_{\ell i}^j (i=1,\dots,m)$ is a clique formed by a subset of vertices of the original polygon $P_\ell^j$. The set $\{V(P_{\ell i}^j): i=1,\dots,m\}$ is mutually disjoint and
\begin{equation}
    V(P_\ell^j) = \bigcup_{i=1}^{m}V(P_{\ell i}^j).
\end{equation}
In the state associated to each new polygon $P_{\ell i}^j$, the probability flux that was going to the vertices in $P_{\ell}^j\backslash P_{\ell i}^j$ should be diverted into the vertices in $P_{\ell i}^j$ so that the state remains unitary, that is,
\begin{equation}
\ket{P_{\ell i}^j} = \sum_{v\in V(P_{\ell i}^j)}\frac{\alpha_{v \ell j}}{\beta_{\ell ij}}\ket{v},
\end{equation}
where
\begin{equation}
    \beta_{\ell ij} = \sqrt{\sum_{v\in V(P_{\ell i}^j)}|\alpha_{v \ell j}|^2}
\end{equation}
is the normalization factor.
The reflection operator associated with tessellation $\mathcal{T}_j$ has now become
\begin{equation}
    U_{\mathcal{T}_j} = 2\left(\sum_{\substack{k=1\\ k\neq \ell}}^{|\mathcal{T}_j|}\ket{P_k^j}\bra{P_k^j}+\sum_{i=1}^m\ket{P_{\ell i}^j}\bra{P_{\ell i}^j}\right)-I.
\end{equation}
The evolution operator of the SQW is given by~(\ref{eq:U}).

Let us see an example for the two-dimensional grid of $4$-cliques (Figure~\ref{fig:4-clique-grid}, $q=1$). Assume that the polygon in position $(0,0)$ is broken into four $1$-clique polygons, as depicted in Figure~\ref{fig:break_poly_green_3}. That is, the polygon which is associated with the quantum state
\begin{equation}
    \ket{\alpha_{00}}=\frac{1}{2}\sum_{k=0}^{3}\ket{x,y,k}
\end{equation}
is broken into the polygons
\begin{equation}
    \ket{\alpha_{00v}} = \ket{v},\quad v\in \{0,1,2,3\}.
\end{equation}
The reflection operator associated with the green tessellation is now described as
\begin{equation}
    U_{0} = 2\left(\sum_{\substack{x,y=0\\ (x,y)\neq (0,0)}}\ket{\alpha_{x,y}}\bra{\alpha_{x,y}}+\sum_{v=0}^{3}\ket{\alpha_{00v}}\bra{\alpha_{00v}}\right)-I.
\end{equation}
The operator associated with the red tessellation is unchanged.

\subsubsection*{Equivalence to the Flip-flop Coined QW model}

Now consider we have the staggered model obtained from the coined model. 
In this case, there are two types of tessellations. One tessellation is associated with the coin action and the other is associated with the shift action. For example, in Fig.~\ref{fig:4-clique-grid}, if we consider that the graph is obtained from a flip-flop coined QW on the two-dimensional grid, the green tessellation is associated with the coin operator and the red tessellation is associated with the shift operator. 

Remember that the polygons of the shift tessellation consist of 2-cliques. The only way to break a polygon in this case is to break it into two 1-clique polygons.
Therefore, breaking a polygon of the shift tessellation will be equivalent to not apply the shift operator to the related edge in the FCQW version. We can think of it as removing an edge in the FCQW, but recall that the coin operator will still be acting as if there is an edge, in this case. 

Breaking polygons of the coin tessellation will be equivalent to modify the coin in the associated vertex of the FCQW version.
For example, if we break the polygon into 1-cliques polygons, then it will be equivalent of not applying the coin operator in the related vertex of the FCQW version (or the coin is the identity). 

\subsection{Breaking vertices}

Now suppose we remove a vertex from the graph. In this case, we have to make arrangements in all the tessellations (since a tessellation covers all vertices in the graph). We should remove the vertex from the polygons which contains it. Since, each polygon is a clique, by removing one of its vertices it will continue to be a clique (not breaking any properties).
Let us suppose vertex $v$ is removed. Then, the probability flux that was going to the removed vertex should be diverted to the other vertices. For every polygon $P_i^j$ that contains vertex $v$, the state associated with the polygon becomes
\begin{equation}
    \ket{P_i^j} = \sum_{w \in V(P_i^j)\backslash\{v\}}\frac{\alpha_{wij}}{\beta_{ij}}\ket{w},
\end{equation}
where $\beta_{ij}$ is the normalization factor so that the state can remain unitary, that is,
\begin{equation}
    \beta_{ij} = \sqrt{\sum_{w\in V(P_i^j)\backslash \{v\}}|\alpha_{wij}|^2}.
\end{equation}

\begin{figure}[!htb]
\centering

\begin{subfigure}[t]{0.47\textwidth}
         \centering
    
    \includegraphics[scale=0.55]{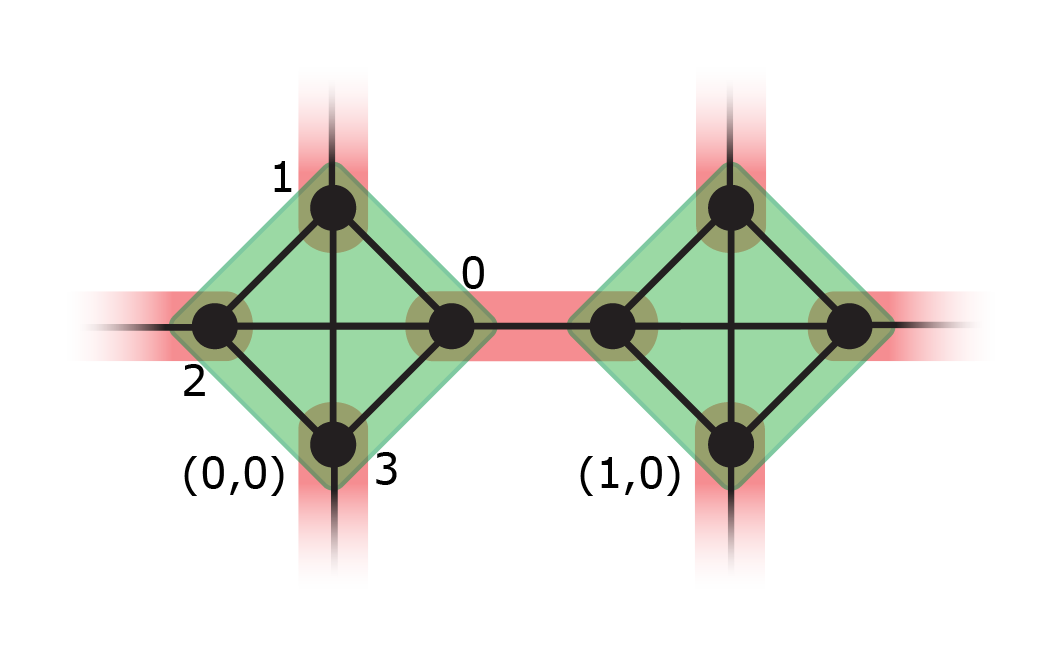}
    \caption{A two-tessellable graph with green and red tessellations.}
    \label{fig:sqw-dec-vertex}
\end{subfigure}
\hspace{5mm}
\begin{subfigure}[t]{0.47\textwidth}
         \centering
\includegraphics[scale=0.55]{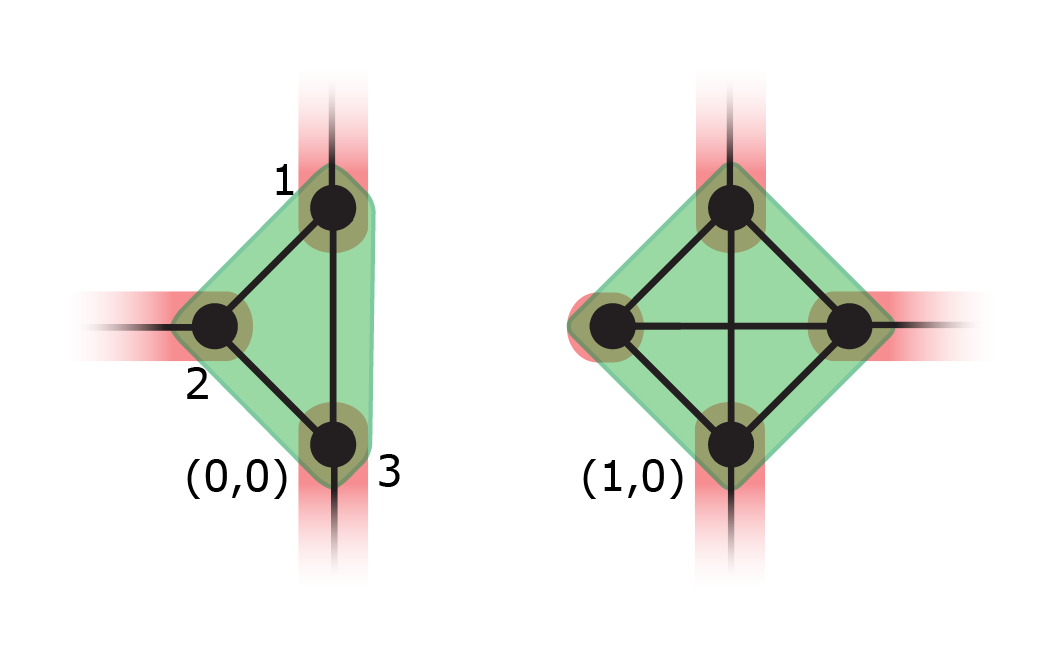}
\caption{Removing vertex $0$ in position $(0,0)$.}
\label{fig:dec-vertex}
\end{subfigure}

\caption{Example of breaking a vertex on the SQW model.}
\label{fig:breaking_vertices}
\end{figure}
For the example shown in Figure~\ref{fig:dec-vertex}, the state associated with the green polygon, which contains the removed vertex $(0,0,0)$,
\begin{equation}
    \ket{\alpha_{00}} = \frac{1}{2}\sum_{k=0}^3\ket{0,0,k}
\end{equation}
becomes
\begin{equation}
    \ket{\alpha_{00}} = \frac{1}{\sqrt{3}}\sum_{k=1}^3\ket{0,0,k}
    \label{eq:alpha000}
\end{equation}
and the state associated with the red polygon, which contains the removed vertex $(0,0,0)$,
\begin{equation}
    \ket{\beta_{00}^{(0)}} = \frac{1}{\sqrt{2}}\left(\ket{0,0,0}+\ket{1,0,2}\right)
\end{equation}
becomes
\begin{equation}
\ket{\beta_{00}^{(0)}} = \ket{1,0,2}.
\label{eq:beta000}
\end{equation}
The operators $U_0$ and $U_1$ associated with each tessellation are obtained as before, considering the new states given by (\ref{eq:alpha000}) and (\ref{eq:beta000}).

\subsubsection*{Equivalence to the Flip-flop Coined QW model}

\begin{figure}[!htb]
\centering
    \begin{subfigure}[t]{0.48\textwidth}
         \centering
         \includegraphics[scale=0.55]{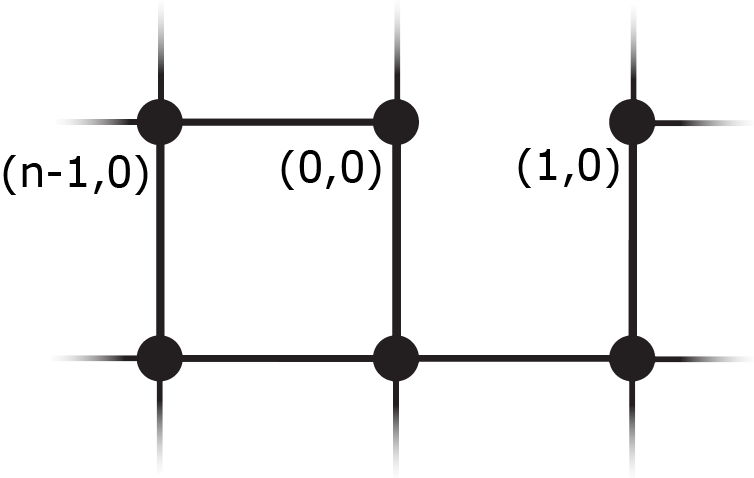}
         \caption{The two dimensional grid with a broken edge.}
         \label{fig:edge-grid}
     \end{subfigure}\hfill
    \begin{subfigure}[t]{0.48\textwidth}
         \centering
         \includegraphics[scale=0.55]{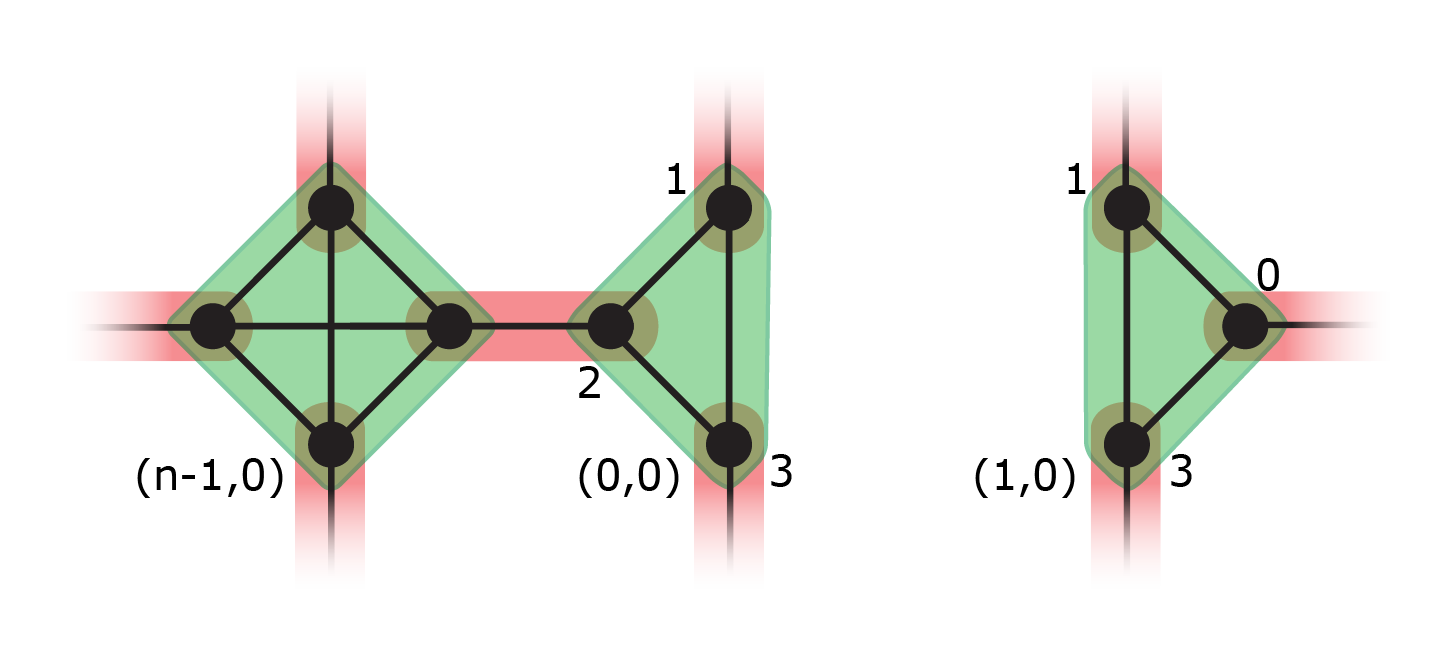}
         \caption{Two vertices removed in the SQW equivalent of the FCQW version.}
         \label{fig:edge-sqw-grid}
     \end{subfigure}
\caption{Example of breaking an edge in the coined version and its equivalent on the staggered model.}
\label{fig:breaking_edge}
\end{figure}

Consider the staggered model obtained from the coined model. Removing a vertex in the SQW model affects the shift and coin tessellations. 
If the removed vertex belongs to a 1-clique polygon of the coin tessellation, then it will be equivalent to removing the associated vertex in the coined model. Otherwise,
it will be equivalent to modifying the action of the coin and shift operators in the vertex of the FCQW associated to the clique which contains the removed vertex. Removing a vertex in the SQW model also affects the shift action on the neighbor vertex of the FCQW associated to its direction.
We can think of it almost as removing an edge in the coined version. Only if we remove the clique which belongs to the polygon in the shift tessellation, then it will be equivalent to removing an edge in the coined version, as depicted in Figure~\ref{fig:breaking_edge}. 

\section{Numerical simulations}\label{sec:simulations}

Consider the SQW on the two-dimensional grid of 4-cliques (Figure~\ref{fig:4-clique-grid}). This graph is the equivalent to the FCQW on the two-dimensional grid. Let the initial state of the SQW be
\begin{equation}
    \ket{\psi(0)} = \frac{1}{2}\sum_{k=0}^3\ket{0,0,k},
    \label{eq:psi0}
\end{equation}
which means we are starting at position $(0,0)$. The probability of obtaining the position $(x,y)$ at time $t$ after measurement in the computational basis is given by
\begin{equation}
    P_t(x,y) = \bra{\psi(t)}\left(\ket{x,y}\bra{x,y}\otimes I_4\right)\ket{\psi(t)},
    \label{eq:pt}
\end{equation}
that is, we sum the probability at each vertex of the clique in position $(x,y)$ from the state at time $t$. 

\begin{figure}[!htb]
\centering
         \centering
         \includegraphics[width=0.8\textwidth]{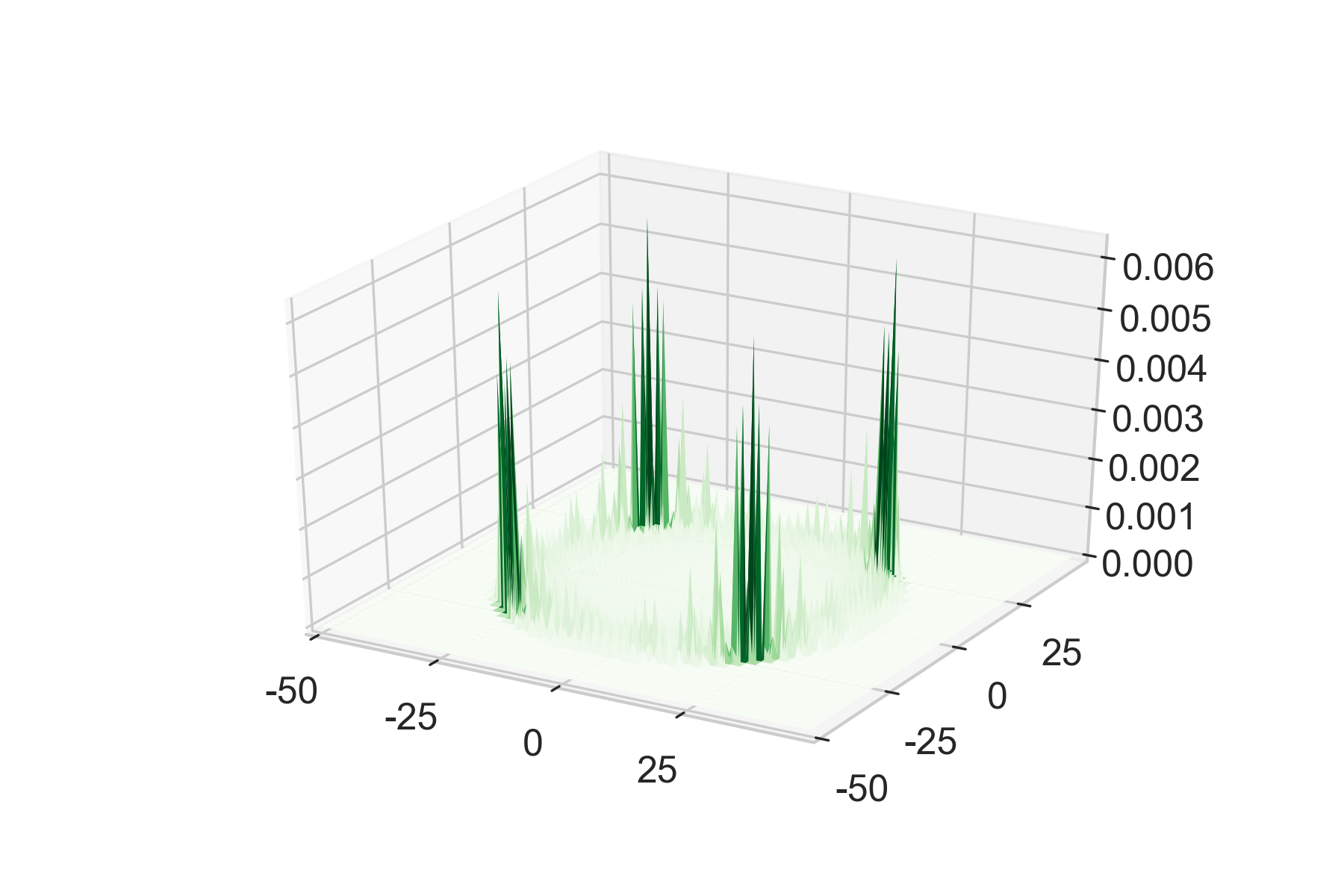}
\caption{Probability distribution  of the two-dimensional grid for the SQW on the two-dimensional grid of $4$-cliques with $n=100$ at $t=50$. The initial state is given by (\ref{eq:psi0}). }
\label{fig:dist-p0}
\end{figure}

From (\ref{eq:pt}), we can obtain the probability distribution of the walk on the two-dimensional grid as depicted in Figure~\ref{fig:dist-p0} for $n=100$, $t=50$ and initial state~(\ref{eq:psi0}). We can observe that the walk is symmetric and delocalized. The same behavior can be observed for the non-flip-flop coined QW with Grover and other coins~\cite{Oliveira:2006}.   

\begin{figure}[!htb]
\centering
    \begin{subfigure}[t]{0.48\textwidth}
         \centering
         \includegraphics[scale=0.5]{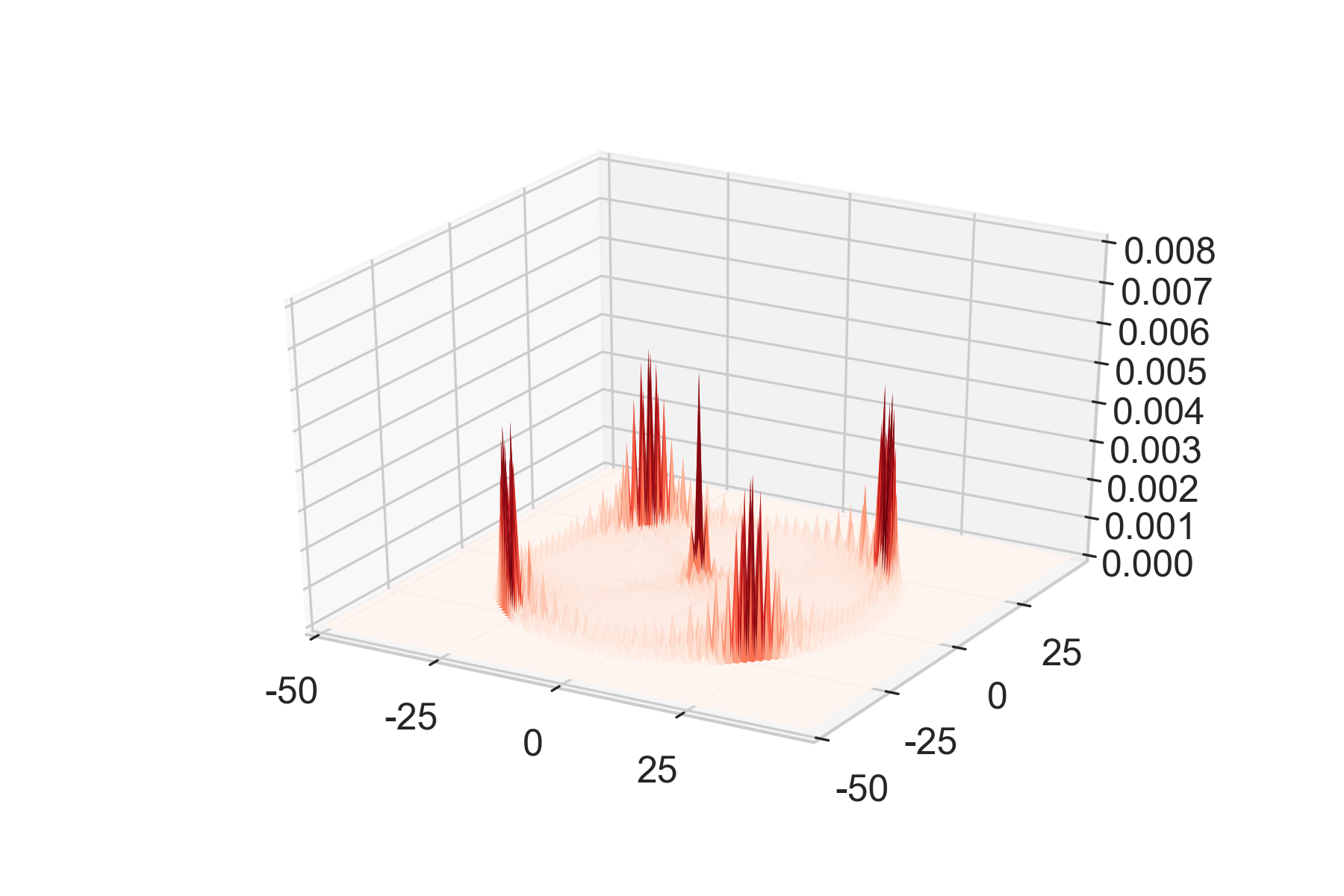}
         \caption{Breaking vertices, $p=0.001$}
         \label{fig:prob_p0001_3d_vert}
     \end{subfigure}\hfill
    \begin{subfigure}[t]{0.48\textwidth}
         \centering
         \includegraphics[scale=0.5]{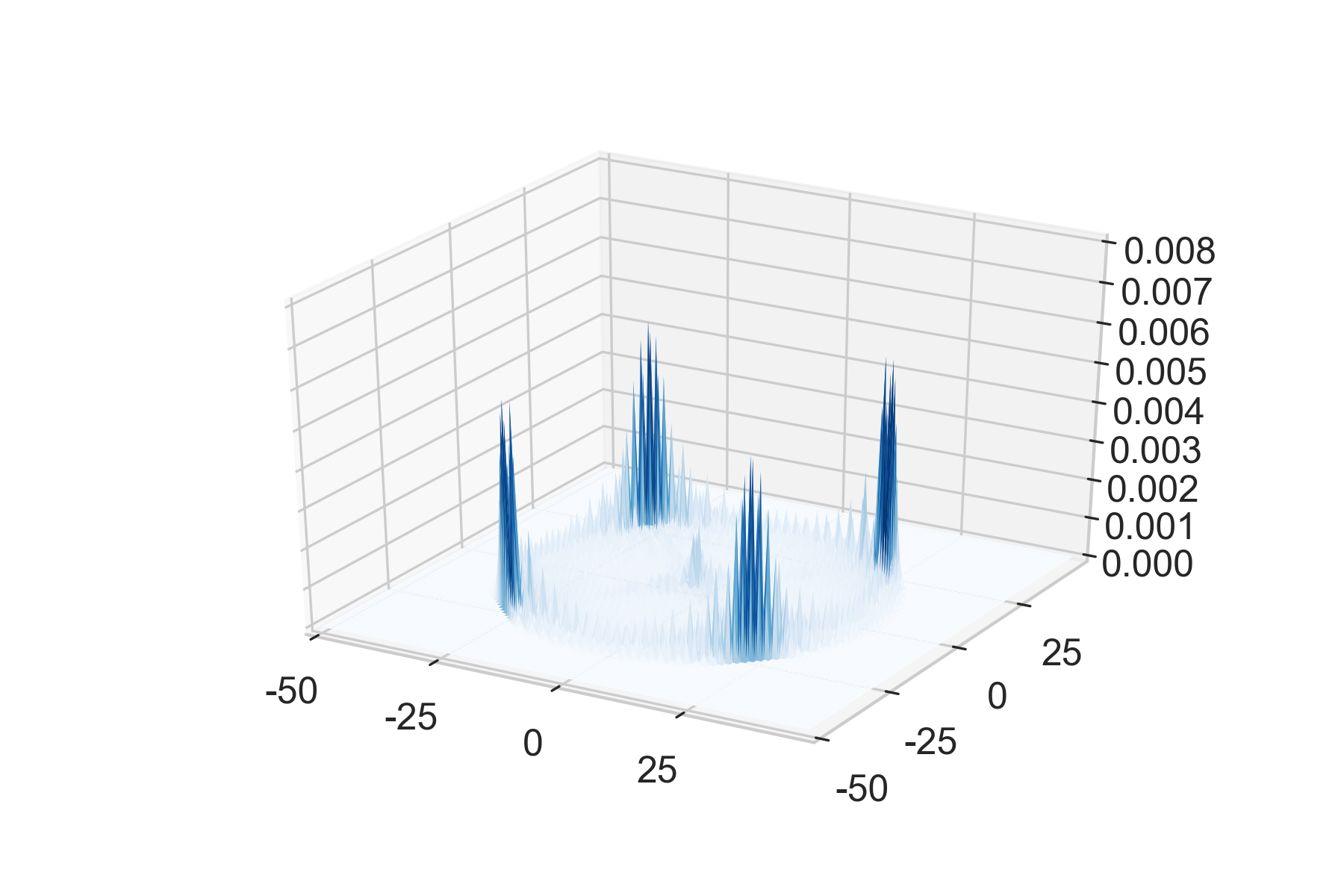}
         \caption{Breaking polygons, $p=0.001$}
         \label{fig:prob_p0001_3d_poly}
     \end{subfigure}\\
     
     \begin{subfigure}[t]{0.48\textwidth}
         \centering
         \includegraphics[scale=0.5]{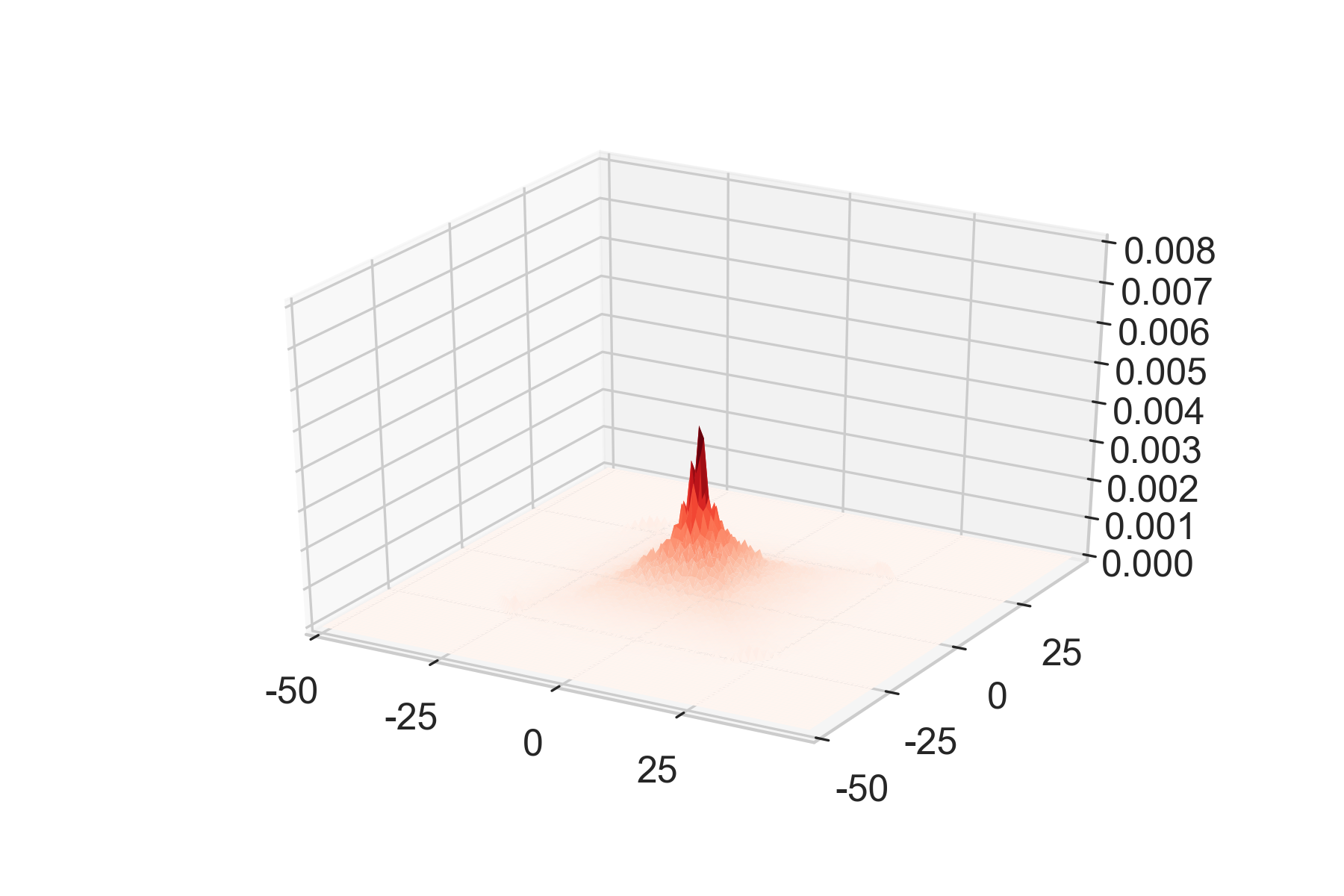}
         \caption{Breaking vertices, $p=0.01$}
         \label{fig:prob_p001_3d_vert}
     \end{subfigure}\hfill
    \begin{subfigure}[t]{0.48\textwidth}
         \centering
         \includegraphics[scale=0.5]{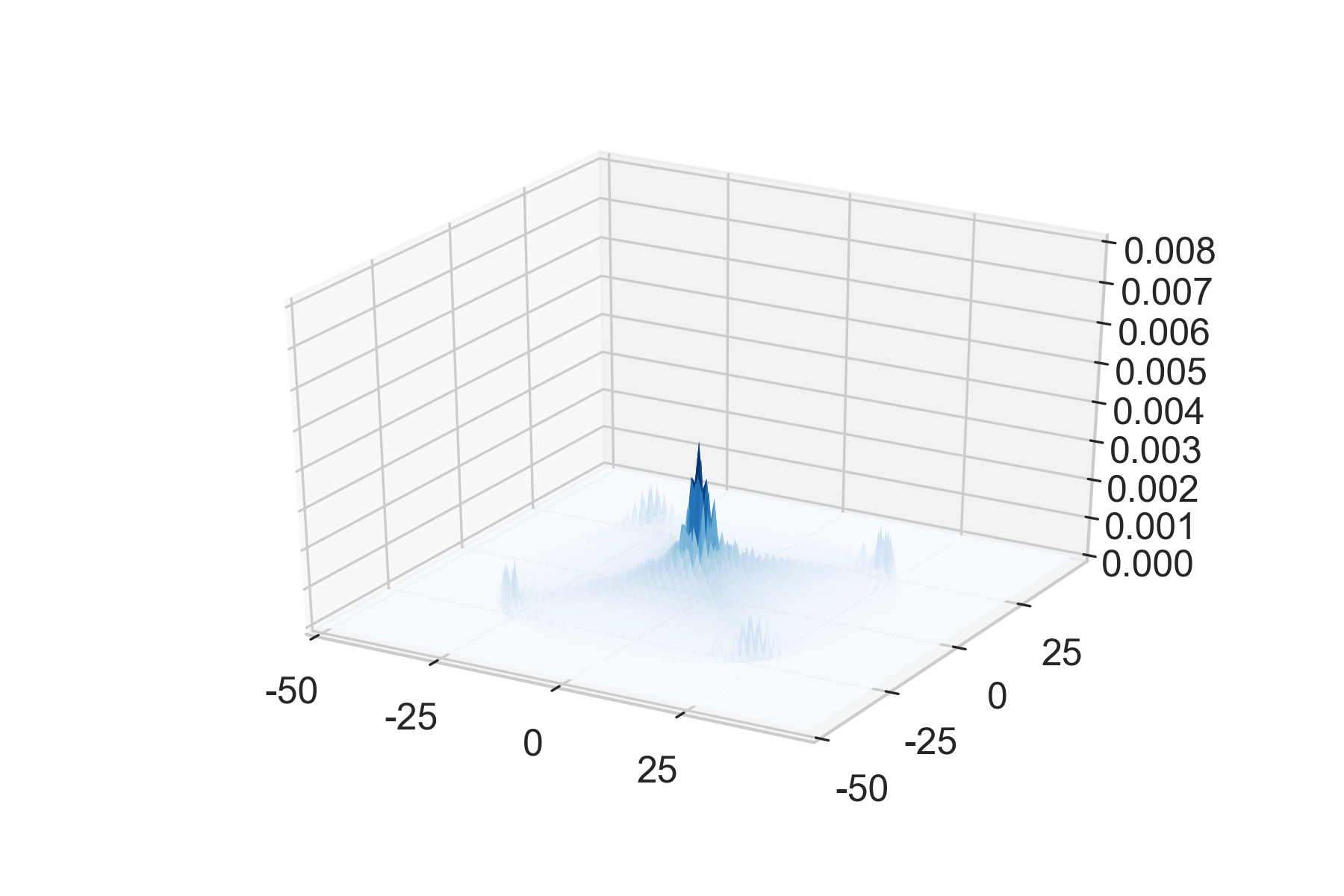}
         \caption{Breaking polygons, $p=0.01$}
         \label{fig:prob_p001_3d_poly}
     \end{subfigure}\\
     
     \begin{subfigure}[t]{0.48\textwidth}
         \centering
         \includegraphics[scale=0.5]{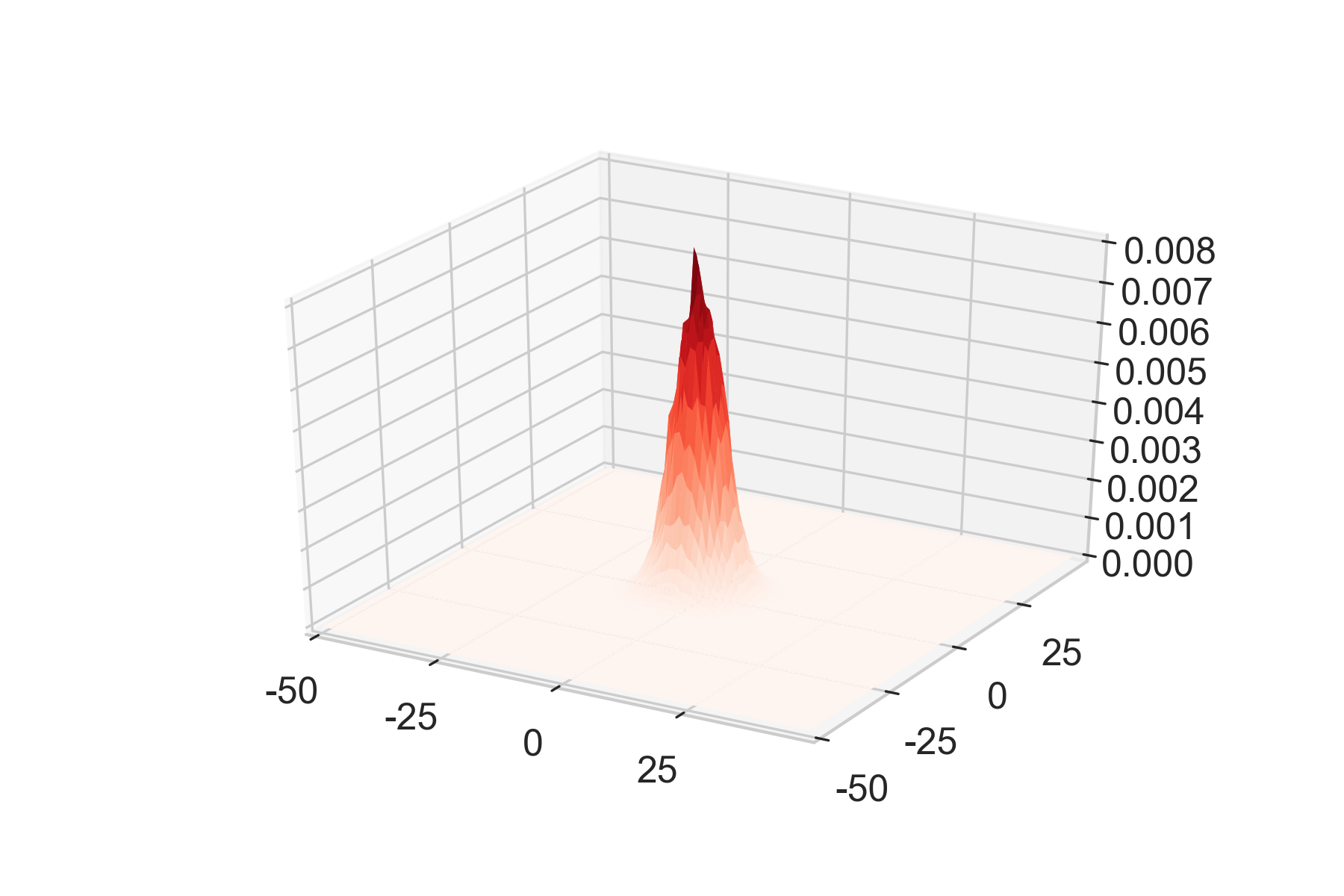}
         \caption{Breaking vertices, $p=0.1$}
         \label{fig:prob_p01_3d_vert}
     \end{subfigure}\hfill
    \begin{subfigure}[t]{0.48\textwidth}
         \centering
         \includegraphics[scale=0.5]{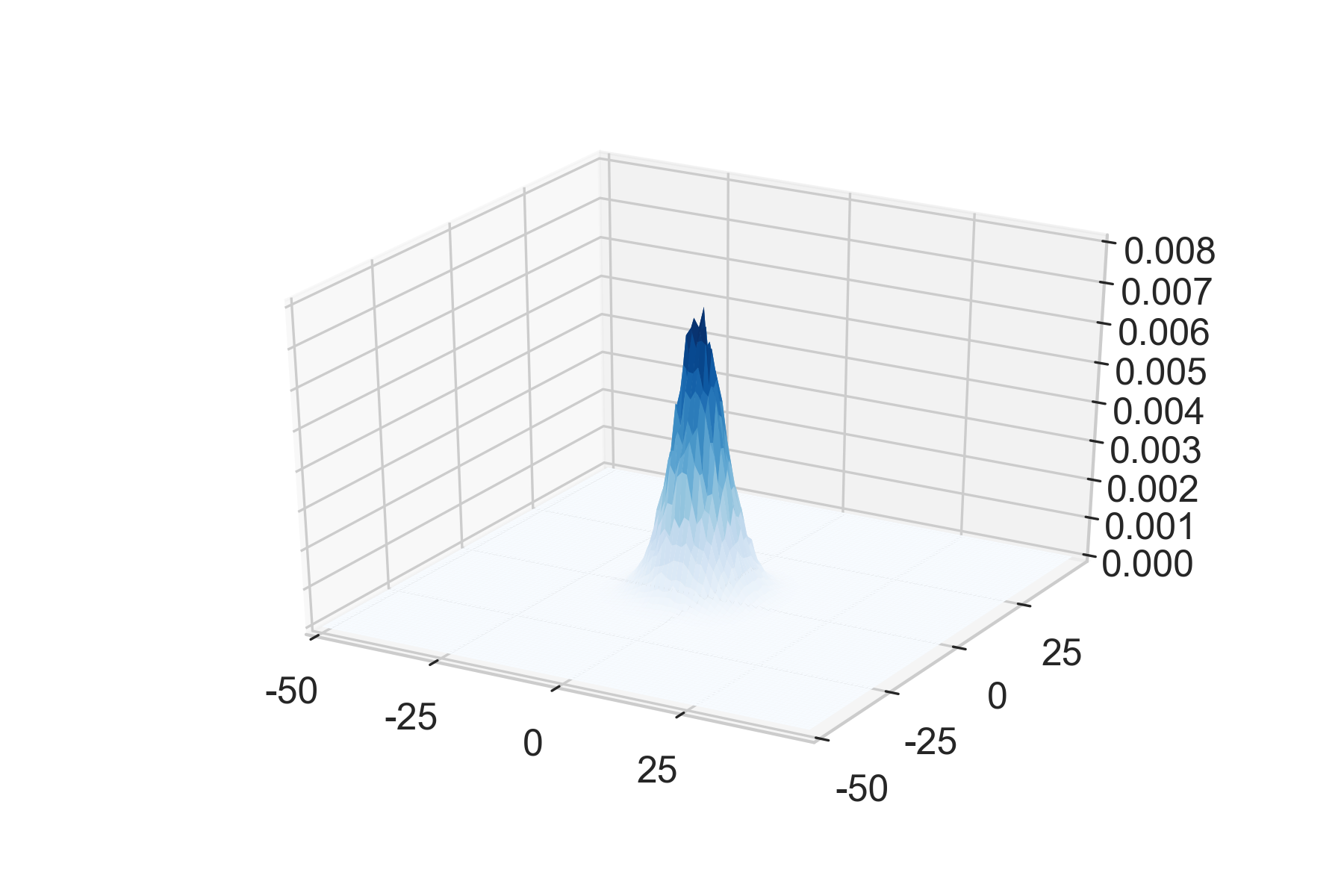}
         \caption{Breaking polygons, $p=0.1$}
         \label{fig:prob_p01_3d_poly}
     \end{subfigure}
\caption{Average of the probability distribution over 100 runs, for the SQW on the two-dimensional grid of 4-cliques with $n=100$. Breaking vertices (a),(c),(e). Breaking polygons (b), (d), (f).}
\label{fig:dist-dec}
\end{figure}

Now we analyze what happens to the SQW when affected by the decoherence models described in Section~\ref{sec:decoherence}. Let $p$ be the probability of breaking each polygon or vertex of the graph before step $t$ of the quantum walk, that is, we consider a model of dynamic percolation. In the case of breaking polygons, we break it into 1-clique polygons. Figure~\ref{fig:dist-dec} shows the behavior of the probability distribution for $p\in\{0.001,0.01,0.1\}$. The quantum behavior of delocalization is still quite present for $p=0.001$ (Figures~\ref{fig:prob_p0001_3d_vert} and~\ref{fig:prob_p0001_3d_poly}). The classical behavior starts to develop as we increase the value of $p$. For the breaking polygons case, it seems the classical behavior emerges a little bit slower than the breaking vertices case.
For $p=0.1$ both plots (Figures~\ref{fig:prob_p01_3d_vert} and~\ref{fig:prob_p01_3d_poly}) present a fully developed classical behavior.

\begin{figure}[!htb]
\centering
    \begin{subfigure}[t]{0.48\textwidth}
         \centering
         \includegraphics[scale=0.5]{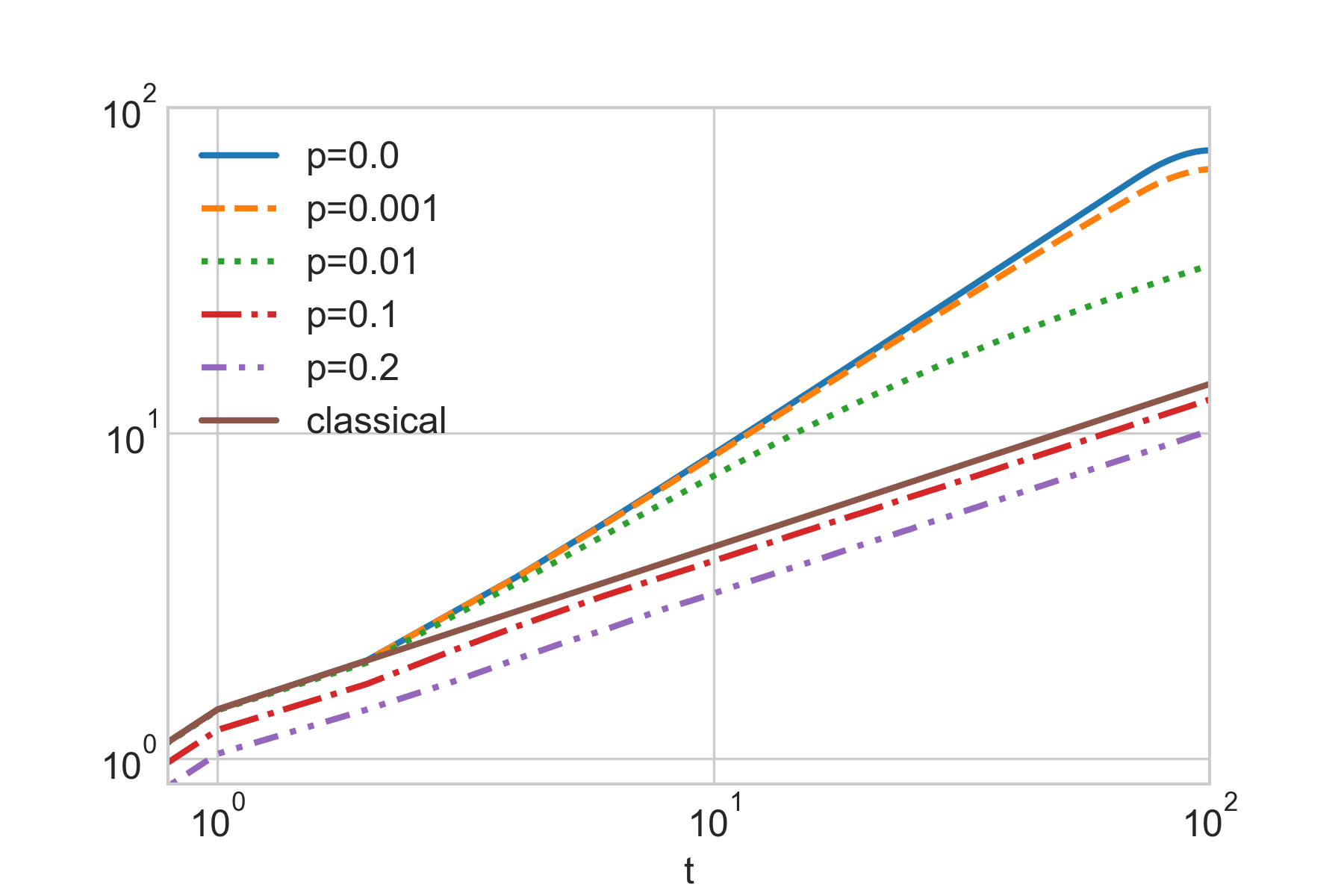}
         \caption{Breaking vertices}
         \label{fig:std_vert}
     \end{subfigure}\hfill
    \begin{subfigure}[t]{0.48\textwidth}
         \centering
         \includegraphics[scale=0.5]{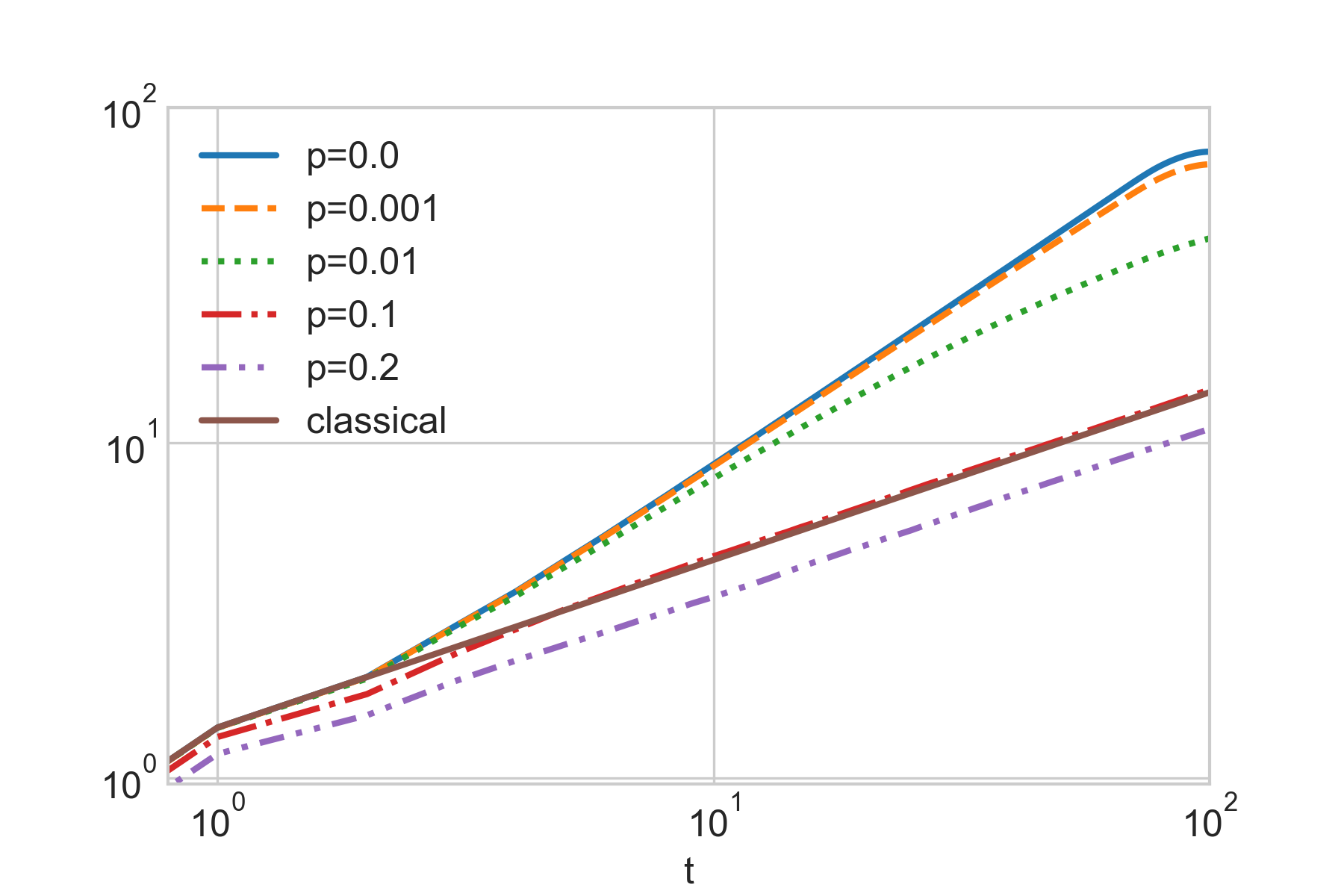}
         \caption{Breaking polygons}
         \label{fig:std_poly}
     \end{subfigure}
\caption{Standard deviation for the SQW on the two-dimensional grid of 4-cliques with $n=100$ considering the decoherence model of (a) breaking vertices and (b) breaking polygons, in the cases $p=\{0.0,0.001,0.01,0.1,0.2\}$. The solid brown curve depicts the classical case on the two-dimensional grid.}
\label{fig:std}
\end{figure}

The transition from the quantum to the classical behavior can be better observed by the plot of the standard deviation $\sigma$, as depicted in Figure~\ref{fig:std} for $n=100$. Figure~\ref{fig:std_vert} shows the standard deviation for the breaking vertices case and figure~\ref{fig:std_poly}, for the breaking polygons case. The solid blue curve is the quantum case without decoherence ($p=0$). Notice that there is an inclination of the curve when reaching $t=100$ which happens because we have periodic bounded conditions of the grid. The solid brown curve shows the classical case. We can observe how the walk lose its quantum behavior when increasing the value of $p$. A similar behavior can be seen for the coined QW with the Grover coin on the two-dimensional grid with broken links~\cite{Oliveira:2006}.

\subsection{How decoherence affects search}

Suppose we are searching for a missing green polygon on the two-dimensional grid of 4-cliques. This is equivalent to search for a marked vertex in the two-dimensional grid using the FCQW model, as mentioned before.
We analyze the decoherence models, described in Section~\ref{sec:decoherence}. Let $p$ be the probability of breaking each polygon or  vertex of the graph before step $t$ of the quantum walk. In the case of breaking polygons, we break it into 1-clique polygons. 
We run the simulations for $p \in \{0.001,0.01,0.1\}$ until the maximum number of steps is achieved. According to some simulations, the maximum number of steps of $1.5\sqrt{N\log N}$ is enough in order to obtain the maximum peak of the success probability. We obtain the average over 100 runs for each value of $p$. 
The results are compared with the simulation without decoherence ($p=0$). 

\begin{figure}[!htb]
\centering
\begin{subfigure}[t]{0.48\textwidth}
         \centering
\includegraphics[scale=0.5]{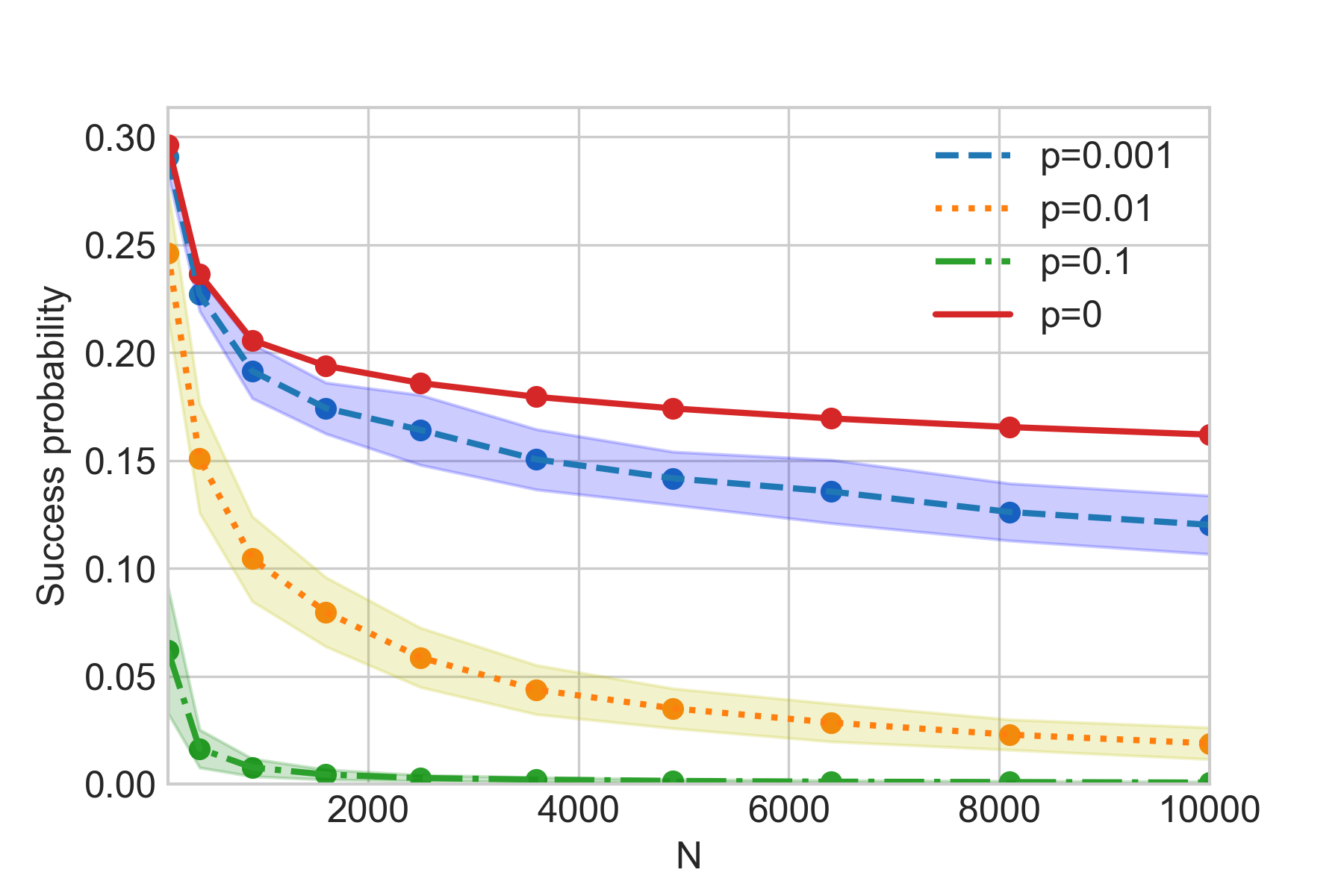}
\caption{Breaking vertices.}
\label{fig:dec_vertices_prob}
\end{subfigure}
\hfill
\begin{subfigure}[t]{0.48\textwidth}
         \centering
\includegraphics[scale=0.5]{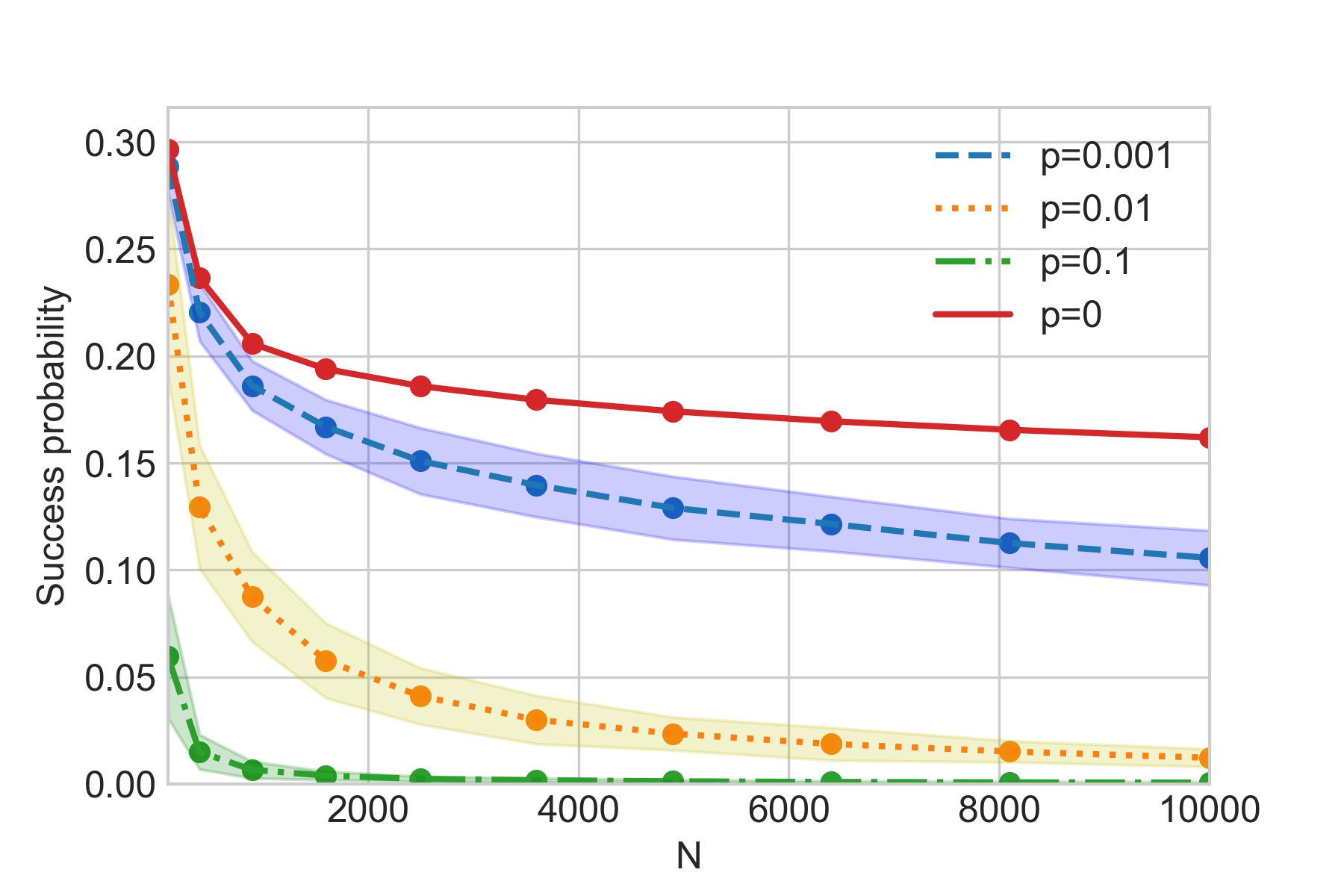}
\caption{Breaking polygons into 1-clique polygons.}
\label{fig:dec_polygons_prob}~\qquad
\end{subfigure}
\caption{The average success probability for $n = 10, 20, \dots, 100$ ($N = n^2$). The solid red curve is for $p=0$; the dashed blue for $p=0.001$; the dotted orange for $p=0.01$; the dot-dashed green for $p=0.1$. The light area around the curves shows the confidence bands.}
\label{fig:dec_prob}
\end{figure}

Figure~\ref{fig:dec_prob} shows the average probability of finding the marked vertex, that is, the success probability, for $n=10, 20, \dots,100$. The case of breaking vertices is depicted in Fig.~\ref{fig:dec_vertices_prob} and the case of breaking polygons, in Fig.~\ref{fig:dec_polygons_prob}.  The solid red curve represents $p=0$; the dashed blue, $p=0.001$; the dotted orange, $p=0.01$; and the dot-dashed green, $p=0.1$. The confidence bands for each simulated probability is depicted by the light area around each curve. As expected, the value of the success probability decreases as the value of $p$ increases. All curves for $p \in \{0.001,0.01,0.1\}$ are below the curve for $p=0$.

\begin{figure}[!htb]
\centering
\begin{subfigure}[t]{0.48\textwidth}
         \centering
\includegraphics[scale=0.5]{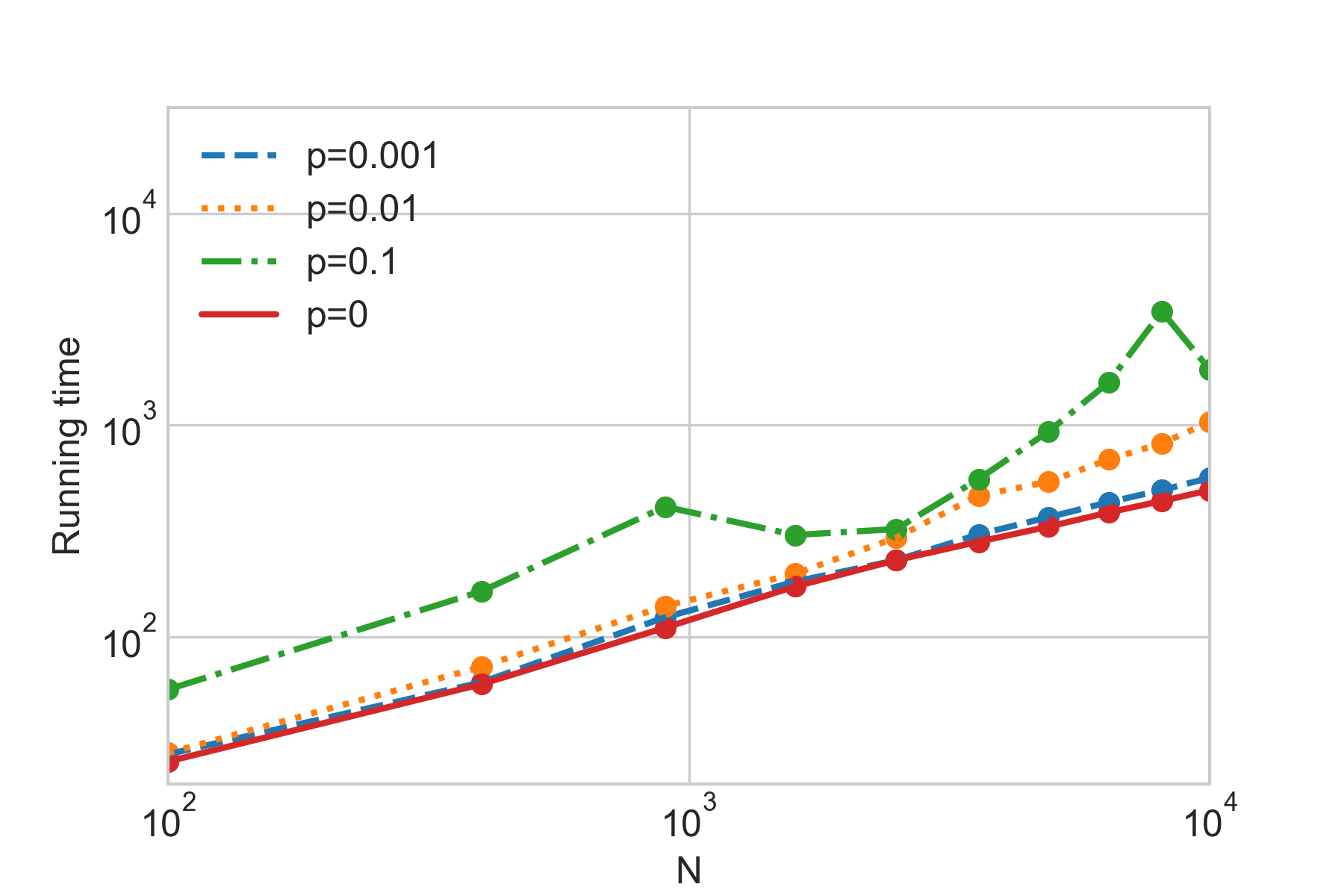}
\caption{Breaking vertices.}
\label{fig:dec_vertices_time}
\end{subfigure}\hfill
\begin{subfigure}[t]{0.48\textwidth}
         \centering
\includegraphics[scale=0.5]{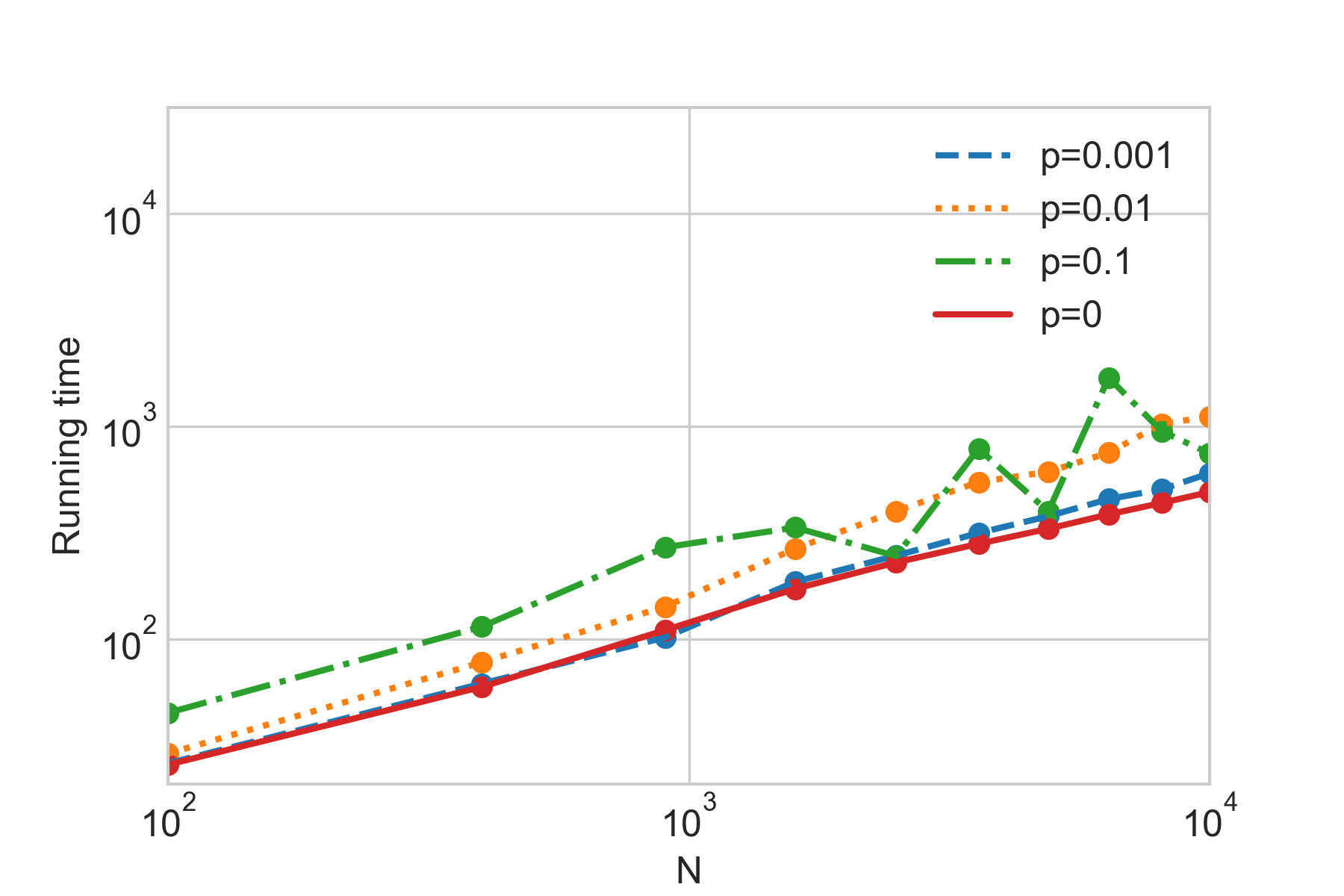}
\caption{Breaking polygons into 1-clique polygons.}
\label{fig:dec_polygons_time}
\end{subfigure}
\caption{The average running time for $n = 10, 20, \dots, 100$ ($N = n^2$) in logarithmic scale. The solid red curve is for $p=0$; the dashed blue for $p=0.001$; the dotted orange for $p=0.01$; the dot-dashed green for $p=0.1$.}
\label{fig:dec_running_time}
\end{figure}

Figure~\ref{fig:dec_running_time} shows the average running time, \emph{i.e.}, the number of steps to reach the maximum of the average success probability divided by the square root of the average success probability. The breaking vertices and breaking polygons cases are depicted in Figs.~\ref{fig:dec_vertices_time} and \ref{fig:dec_polygons_time}, respectively. The dashed blue curve ($p=0.001$) is quite close to the case without decoherence. The running time increases by increasing the value of $p$. For $p=0.01$ the curve starts to have a higher inclination than the case for $p=0$. The behavior for $p=0.1$ is already quite erratic since for this value of probability the walk has already developed its classical behavior, as shown before.  

\subsubsection{Expanding the tessellations intersection}

As mentioned before, searching for a clique on the green polygons in any of the two-dimensional grids of $4q$-cliques is equivalent~\cite{Santos:2019}. Let us analyze how robust against decoherence can the SQW be as we increase the value of $q$.
We analyze the decoherence model of breaking polygons. For this case, we consider that each polygon has probability $p$ of being broken into two polygons -- one with $1$-clique and the other containing a $(4q-1)$-clique. The vertices in each of the two polygons are chosen at random. We run the simulations for $p \in \{0.001,0.01\}$ and $q\in \{1,2,3\}$. We obtain the average over 100 runs.

\begin{figure}[!htb]
\centering
\begin{subfigure}[t]{0.48\textwidth}
         \centering
\includegraphics[scale=0.5]{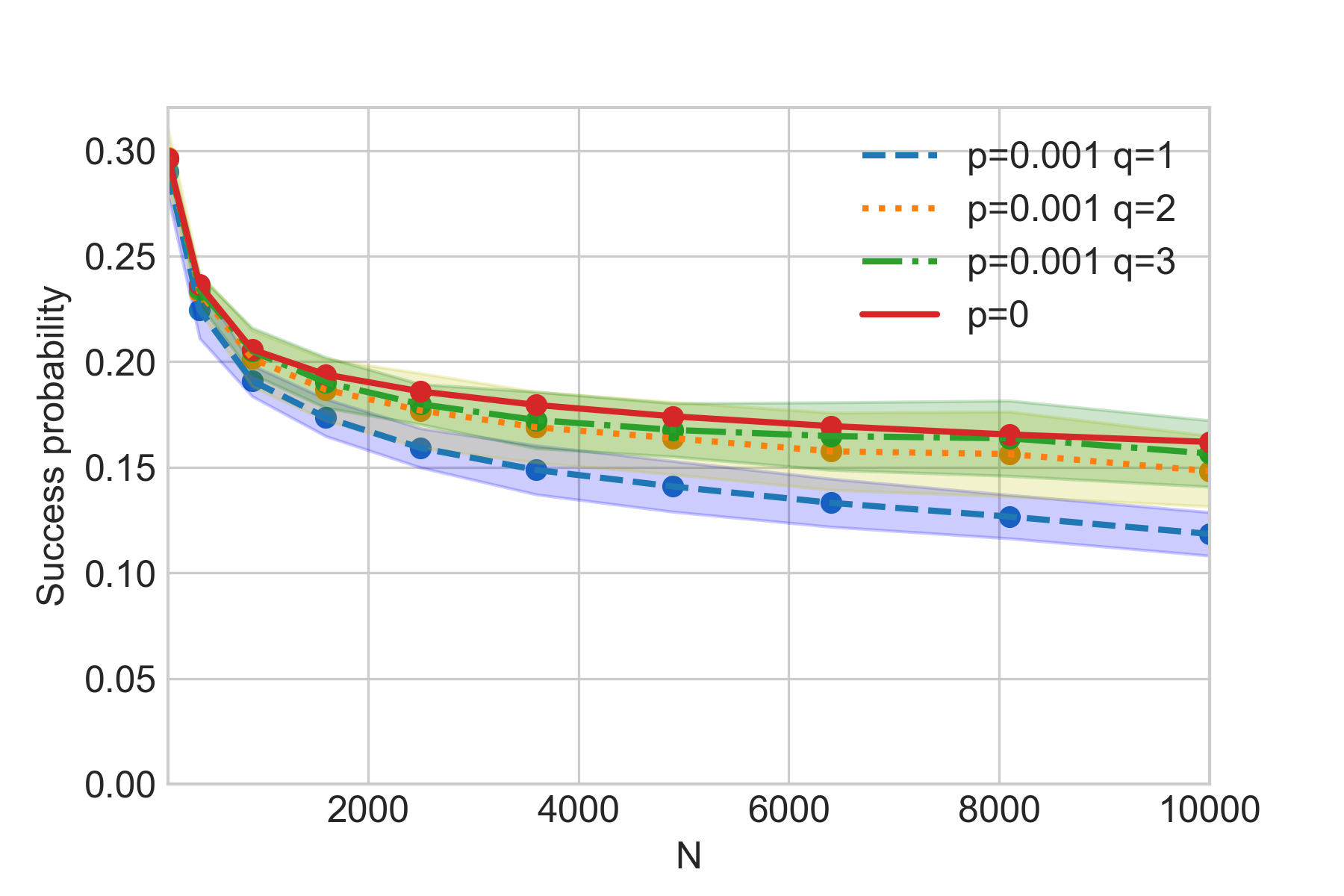}
\caption{$p=0.001$}
\label{fig:dec_vertices_q_prob1}
\end{subfigure}\hfill
\begin{subfigure}[t]{0.48\textwidth}
         \centering
\includegraphics[scale=0.5]{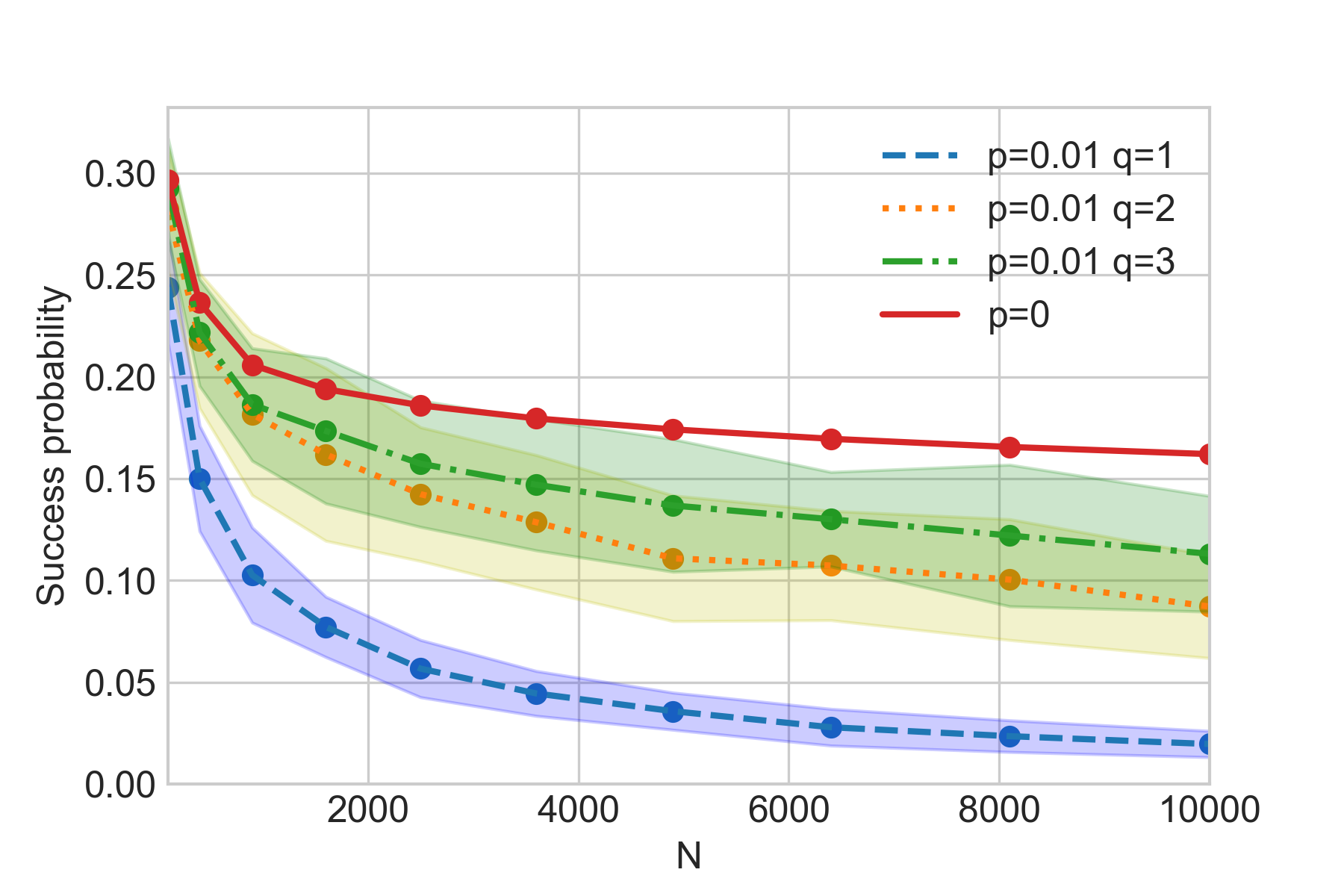}
\caption{$p=0.01$}
\label{fig:dec_vertices_q_prob2}
\end{subfigure}
\caption{The average success probability for $n = 10, 20, \dots, 100$ ($N = n^2$). The dashed blue curve is for $q=1$; the dotted orange curve is for $q=2$; the dot-dashed green curve is for $q=3$; the solid red line if for $p=0$. (a) The case for $p=0.001$. (b) The case for $p=0.01$.  }
\label{fig:dec_prob_expansion}
\end{figure}

Figure~\ref{fig:dec_prob_expansion} shows the average probability of finding the marked vertex, that is, the success probability, for $n=10, 20, \dots,100$. The case for $p=0.001$ is depicted in Fig.~\ref{fig:dec_vertices_q_prob1} and the case for $p=0.01$, in Fig.~\ref{fig:dec_vertices_q_prob2}.   The dashed blue curve represents $q=1$; the dotted orange, $q=2$; and the dot-dashed green, $q=3$. The confidence bands for each simulated probability is depicted by the light area around each curve. We can compare them with the solid red curve for $p=0$.
The value of the success probability is closer to the red curve as $q$ increases.

\begin{figure}[!htb]
\centering
\begin{subfigure}[t]{0.48\textwidth}
         \centering
\includegraphics[scale=0.5]{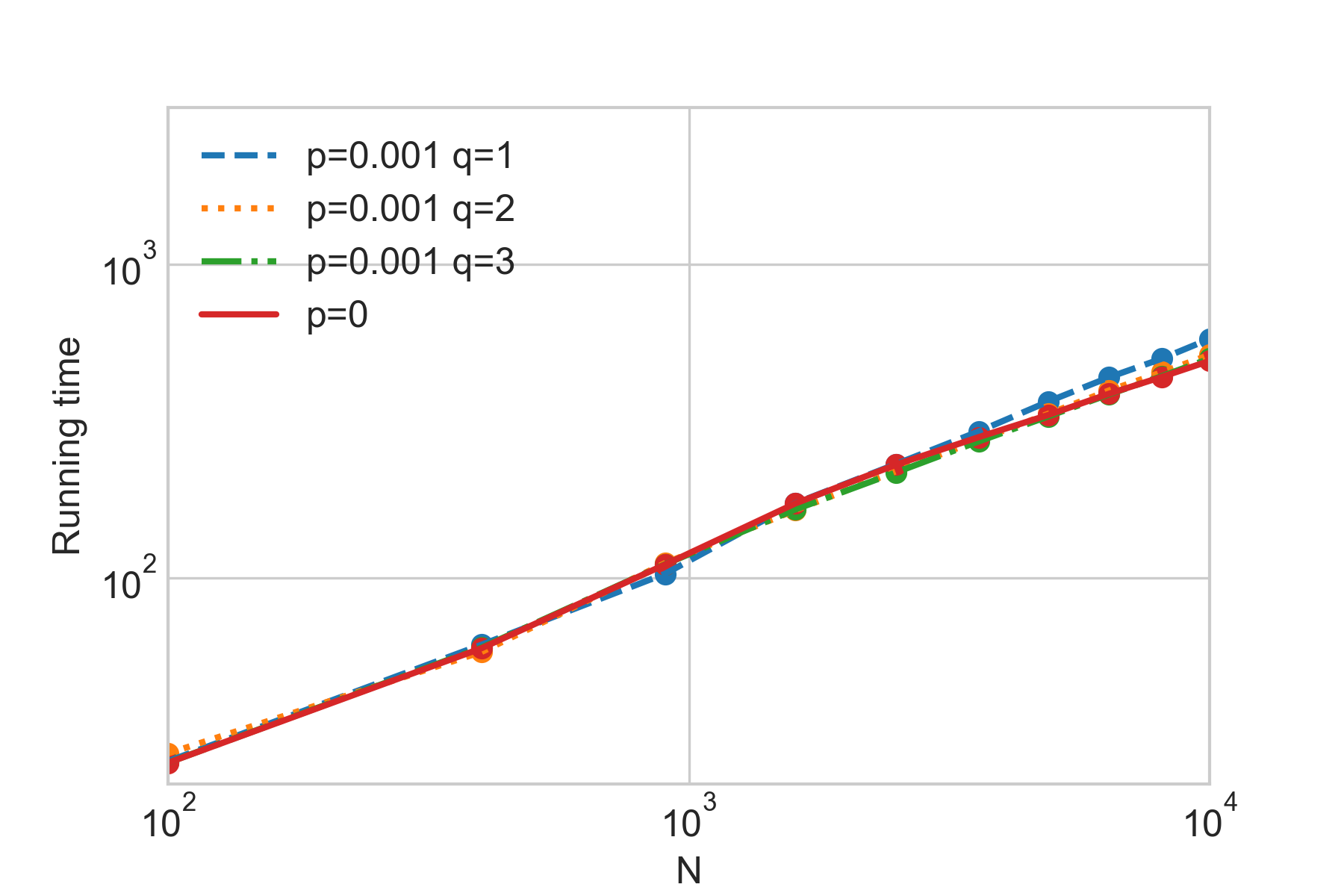}
\caption{$p=0.001$}
\label{fig:dec_vertices_q_time1}
\end{subfigure}\hfill
\begin{subfigure}[t]{0.48\textwidth}
         \centering
\includegraphics[scale=0.5]{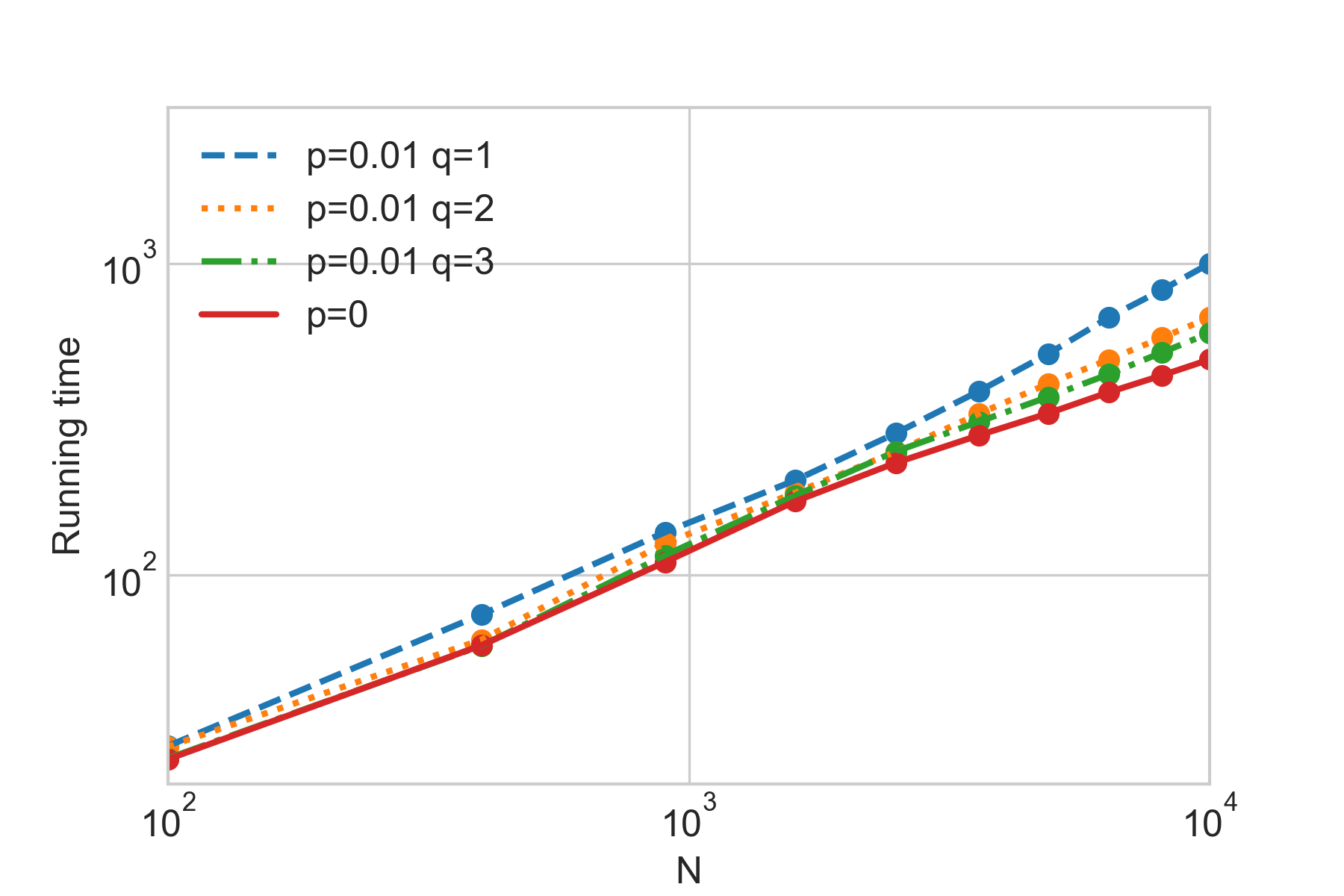}
\caption{$p=0.01$}
\label{fig:dec_vertices_q_time2}
\end{subfigure}
\caption{The average running time for $n = 10, 20, \dots, 100$ ($N = n^2$). The dashed blue curve is for $q=1$; the dotted orange curve is for $q=2$; the dot-dashed green curve is for $q=3$; the solid red curve is for $p=0$. (a) The case for $p=0.001$. (b) The case for $p=0.01$.}
\label{fig:dec_time_expansion}
\end{figure}

\begin{figure}[!htb]
\centering
\begin{subfigure}[t]{0.48\textwidth}
         \centering
\includegraphics[scale=0.5]{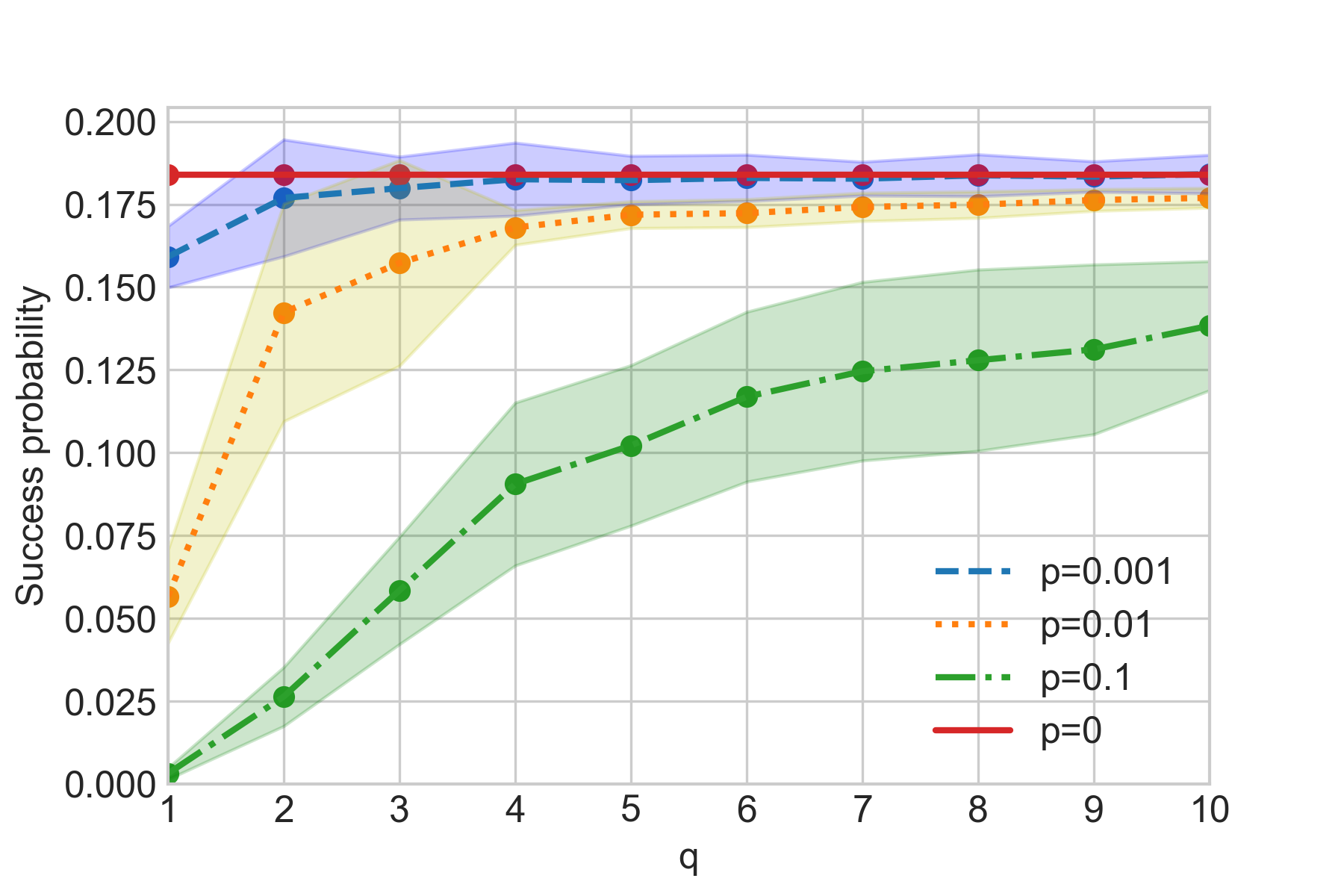}
\caption{Success probability}
\label{fig:dec_qs_prob}
\end{subfigure}\hfill
\begin{subfigure}[t]{0.48\textwidth}
         \centering
\includegraphics[scale=0.5]{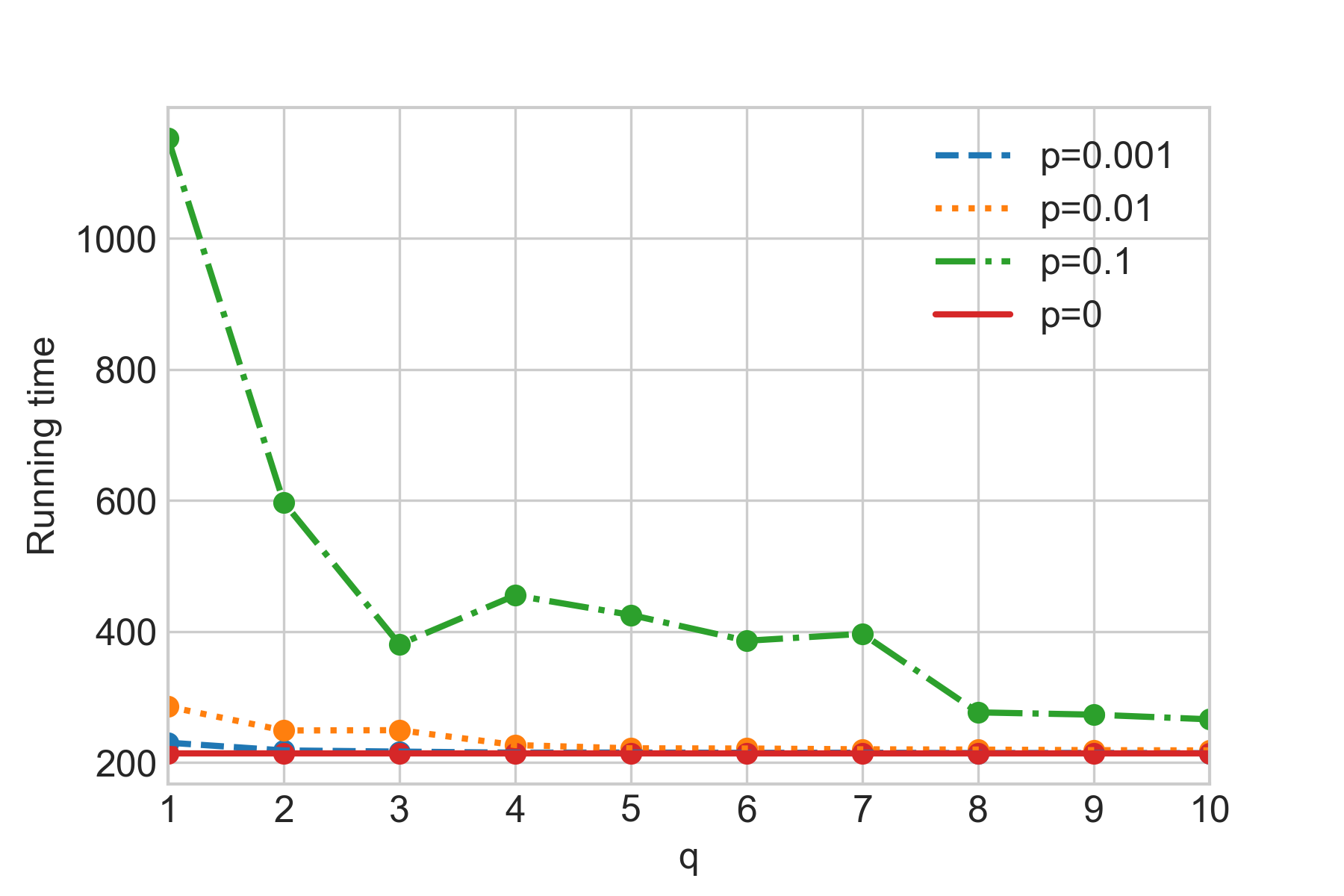}
\caption{Running time}
\label{fig:dec_qs_time}
\end{subfigure}
\caption{The (a) average success probability with confidence bands and (b) average running time for $q = 1,\dots, 10$ and $n=50$. The solid red curve is for $p=0$; the dashed blue is for $p=0.001$; the dotted orange for $p=0.01$; and the dot-dashed green for $p=0.1$.}
\label{fig:dec_qs}
\end{figure}

Figure~\ref{fig:dec_time_expansion} shows the average running time. For $p=0.001$ (Figure~\ref{fig:dec_vertices_q_time1}) the curves coincide with the solid red curve for $p=0$. As $p$ increases we observe that the bigger $q$ the closer to the case without decoherence. This can be seen for $p=0.01$ in Figure~\ref{fig:dec_vertices_q_time2} and in Figure~\ref{fig:dec_qs} where we fix $n=50$ and consider the range $q = 1,\dots,10$. Figure~\ref{fig:dec_qs_prob} and \ref{fig:dec_qs_time} depicts the average success probability and the average running time, respectively, for $p\in\{0.001,0.01,0.1\}$.

\section{Conclusions}\label{sec:conc}

We have analyzed decoherence inspired by percolation on the SQW model. We have presented two models of unitary noise by randomly breaking vertices or polygons of the graph. The evolution operators subject to these noises were obtained. 
The breaking polygons model is equivalent to breaking some edges of the graph, as it allows to break edges in a way that the tessellation number of the graph is not increased. On the other hand, breaking an arbitrary edge of the graph may not be that simple. Each polygon is a clique. By removing an edge, the polygon may not be a clique anymore. In order to fulfill the required properties of the SQW model additional tessellations may be needed. This makes the process non-trivial and  strictly dependent on the structure of the graph.

Additionally, we have shown the equivalence to the coined quantum walk model, when the SQW is obtained from the FCQW model. Breaking vertices can be seen almost as breaking an edge in the coined model. Breaking polygons can be equivalent to modifying the coin or shift operator depending on which tessellation the broken polygon belongs. Therefore, breaking polygons introduces different type of decoherence on the coined model. 

We have numerically analyzed the effect of breaking vertices and polygons on the SQW on the two-dimensional grid of $4$-cliques, which is equivalent to the FCQW on the two-dimensional grid. For both models of decoherence, we have considered a dynamic process where at each time step a vertex or polygon can be broken according to a constant probability $p$. The standard deviation of the quantum walk under decoherence was analyzed for different probabilities and it presents a similar behavior to what happens with the coined quantum walk on the two-dimensional grid with broken links. The quantum walk loses its quantum behavior as we increase the value of the probability. 

The quantum walk based search algorithm under the effect of these noises has also been analyzed on the two-dimensional grid of $4$-cliques. The plots for the breaking vertices and breaking polygons cases are quite similar. The average success probability decreases quite fast as we increase the value of the decoherence probability $p$. When the probability is quite small, say $p=0.001$, the behavior of the decoherent algorithm is still quite close to the algorithm without decoherence. Moreover, we have shown that the algorithm can be more robust against decoherence by expanding the tessellations intersection. For future research, it would be interesting to consider another types of decoherence on SQWs.

\section*{Acknowledgements}
This work was supported by ERDF project number 1.1.1.2/VIAA/1/16/002.


\end{document}